\documentclass[a4paper,12pt]{article}
\usepackage{xcolor}
\usepackage{a4wide}
\usepackage[super]{cite}
\usepackage{mathpazo}
\usepackage{graphicx}
\usepackage[footnotesize, labelfont=bf,it]{caption}
\usepackage{amsmath}
\usepackage{hyperref}
\usepackage[section]{placeins}
\usepackage{booktabs}

\definecolor{RoyalBlue}{cmyk}{1, 0.50, 0, 0}
\hypersetup{colorlinks=true, breaklinks=true, pdfpagemode=UseNone, pageanchor=true, pdfpagemode=UseOutlines,
plainpages=false, bookmarksnumbered, bookmarksopen=true, bookmarksopenlevel=1,
hypertexnames=true, pdfhighlight=/O, filecolor=black,urlcolor=RoyalBlue, linkcolor=black, citecolor=black}

\DeclareMathAlphabet{\pazocal}{OMS}{zplm}{m}{n}

\title{The Scales of Human Mobility}

\author{Laura Alessandretti$^{1,2,\dagger}$, Ulf Aslak$^{1,2,\dagger}$ \& Sune Lehmann$^{1,2,*}$}

\begin{document}

\maketitle

$^1$ Copenhagen Center for Social Data Science, University of Copenhagen, \O ster Farimagsgade 5, Copenhagen K, 1353, Denmark\\
$^2$ Technical University of Denmark, Anker Engelundsvej 1, Kgs Lyngby, DK-2800, Denmark\\
$^\dagger$ These authors contributed equally to this work. \\
$^*$ To whom correspondence should be addressed, sljo@dtu.dk.

\section*{Abstract}
\textbf{%
There is a contradiction at the heart of our current understanding of individual and collective mobility patterns. 
On one hand, a highly influential stream of literature on human mobility driven by analyses of massive empirical datasets finds that human movements show no evidence of characteristic spatial scales. There, human mobility is described as \textit{scale-free}.\cite{brockmann2006scaling, gonzalez2008understanding, song2010modelling} On the other hand, in geography, the concept of \emph{scale}, referring to meaningful levels of description from individual buildings through neighborhoods, cities, regions, and countries, is central for the description of various aspects of human behavior such as socio-economic interactions, or political and cultural dynamics.\cite{paasi2004place, marston2000social} Here, we resolve this apparent paradox by showing that day-to-day human mobility does indeed contain meaningful scales, corresponding to spatial containers restricting mobility behavior. The scale-free results arise from aggregating displacements across containers. We present a simple model, which given a person's trajectory, infers their neighborhoods, cities, and so on, as well as the sizes of these geographical containers.
We find that the containers characterizing the trajectories of more than 700\,000 individuals do indeed have typical sizes. 
We show that our model generates highly realistic trajectories without overfitting and provides a new lens through which to understand the differences in mobility behaviour across countries, gender groups, and urban-rural areas.}

It is nearly impossible to underestimate the importance of establishing a quantitative foundation for our understanding of how individuals move from place to place in their everyday lives. 
Hundreds of millions of individuals spend billions of collective hours commuting every day.\cite{cresswell2006move}
Goods and food are transported through a global network using shared infrastructure.\cite{kaluza2010complex} Understanding mobility patterns helps us mitigate epidemic spreading\cite{kraemer2020effect}, assist in crisis management\cite{song2014prediction}, prepare for dramatic shifts in modes of transportation\cite{becker2017literature}, and many other cases.\cite{cresswell2006move}
For this reason, understanding the origin of scale-free distributions of displacements in empirical mobility traces is crucial, as this issue currently separates the large-scale data-driven human mobility research\cite{barbosa2018human} from the community of human geography\cite{ paasi2004place, marston2000social} and transportation research.\cite{larsen2016mobilities}

Our mental representation of physical space has a hierarchical structure.\cite{hirtle1985evidence}
We describe space referring to \emph{places}\cite{paasi2004place}, meaningful spatial entities with associated typical size, or \textit{scale}, from rooms and buildings -- via neighborhoods, cities, and states -- to nations and continents that are organized in a nested structure.\cite{paasi2004place,von1910isolierte,christaller1980zentralen,berry1967geography,alonso1964location}
Geographical borders confine residential mobility\cite{cadwallader1992migration} and collective mobility fluxes.\cite{thiemann2010structure} Commuting is characterized by a typical travel-time budget, and, as a consequence, there exist characteristic spatial scales that have evolved in connection to the progress of transportation.\cite{marchetti1994anthropological} Further, it has been conjectured that there are fundamental differences between forms of moving at different scales, from moving within a building to traveling across the globe.\cite{berry1967geography, cresswell2006move, larsen2016mobilities}

However, recent empirical research in the field of Human Mobility\cite{barbosa2018human}, has found no evidence for characteristic spatial scales in how people travel.\cite{brockmann2006scaling, song2010modelling, gonzalez2008understanding, noulas2012tale} On the contrary, studies have shown that the distribution of displacement lengths $\Delta r$ travelled by an individual has a power law tail $P(\Delta r) \sim \Delta r^ {-\beta}$ over several orders of magnitude, where typically $1 \leq \beta \leq 2$.\cite{alessandretti2017multi} Power law distributions are also called \textit{scale free}, because they are the only mathematical distribution to have no associated typical scale\cite{newman2005power} (see Supplementary Note~1).

\textbf{Nested scales generate power laws}.
So the question becomes: 
How is it possible that our intuitive conception of space is clearly hierarchical and characterized by typical scales, when a broad range of empirical datasets, ranging from displacements of dollar bills\cite{brockmann2006scaling} or cell-tower data\cite{gonzalez2008understanding} to public transportation systems\cite{liang2013unraveling}, and GPS data\cite{alessandretti2018evidence,gallotti2016stochastic} all suggest that human mobility is scale free?

\begin{figure}[htb!]
\includegraphics[width=\textwidth]{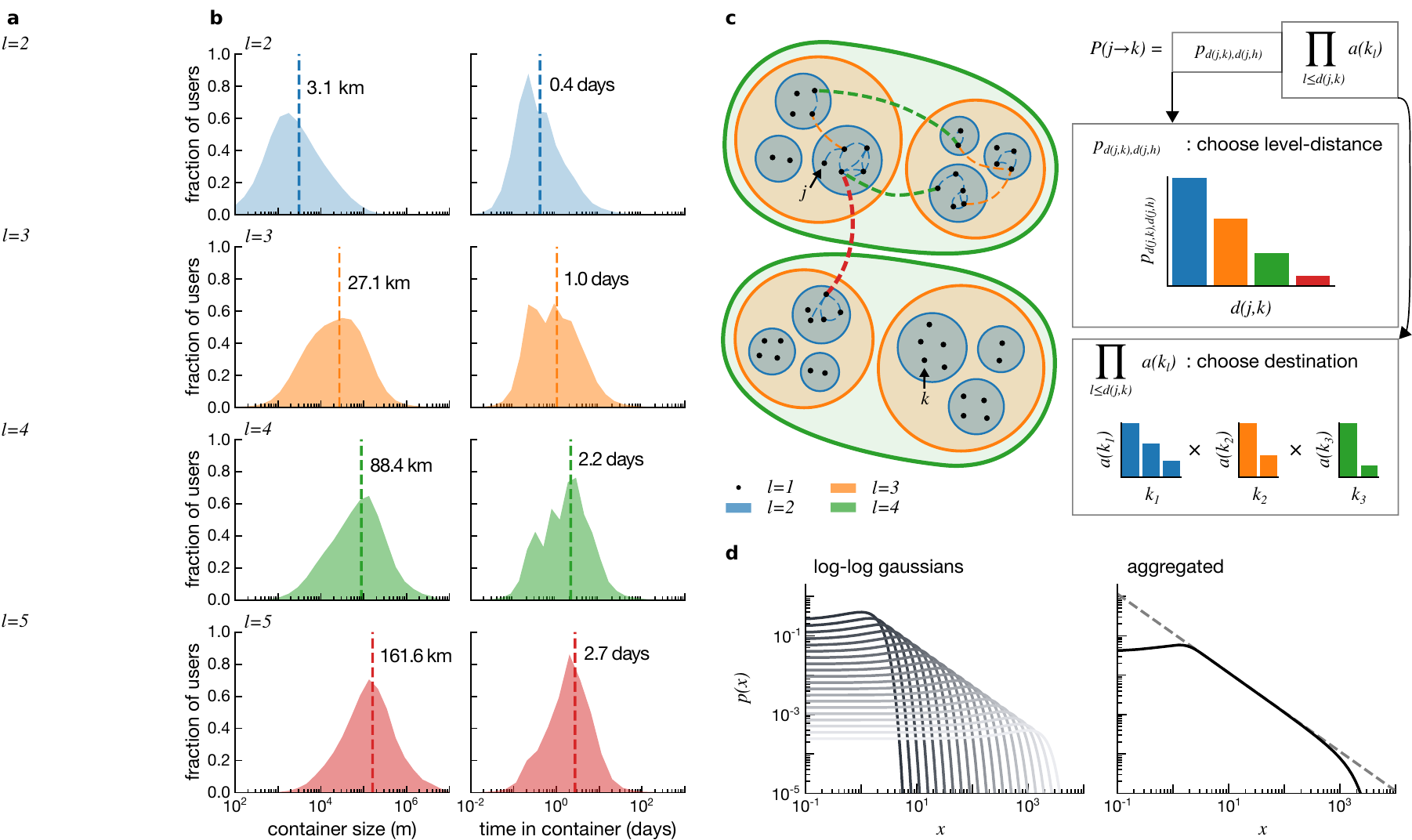}
\caption{\textbf{The Scales of Human Mobility.} a) Example of containers for an individual living in Copenhagen, characterized by the size of containers in neighborhoods (blue), cities (orange), urban agglomerations (green) and regions (red). Map data copyrighted OpenStreetMap contributors and available from \href{https://www.openstreetmap.org}{https://www.openstreetmap.org}. b) Distribution of container sizes (left column) and median time spent in the same container (right column) across individuals. Dashed lines correspond to medians. Results, shown here for containers at different hierarchical levels, are obtained by fitting the \emph{container model} to the D1 dataset, consisting of $\sim700\,000$ anonymized GPS traces of individuals distributed across the globe (see Extended Data Figure~\ref{Extended_data_F2} for dataset D2). c) Schematic representation of the container model. Individuals move between locations (black dots) inside a nested set of containers. The probability of transitioning between two locations $j$ and $k$ is the product of two factors, corresponding to choosing level-distance and destination (see main text). d) Gaussian distributions with different variances (left panel) and their mixture (right panel) in log-log scale. The dashed line (right panel) is a power law $P(x)\sim x^{-\beta}$ with $\beta=1$ to guide the eye. \label{Figure1}}
\end{figure}

To explain this apparent contradiction, we propose that each typical scale of human mobility corresponds to a \textit{container} of a certain mobility behavior. 
These containers (rooms, buildings, neighborhoods, cities, countries, and so on) have typical sizes, see Figure~\ref{Figure1}a, and roughly correspond to the notion of \textit{places} in geography.\cite{paasi2004place} The observed power law arises when we aggregate mobility behavior within containers and mobility that transports a person between containers.
Specifically, it is well known that mixtures of normal (or lognormal) distributions with different variances can generate power laws\cite{gheorghiu2004heterogeneity} (see Figure~\ref{Figure1}d). 
More specifically, we assume that for each individual, physical space is organized as a nested structure of containers. This structure relates, in part, to the organization of the transportation system\cite{marchetti1994anthropological} and to the concrete structure of our built environment\cite{berry1967geography}, see Figure~\ref{Figure1}a.

We propose that these nested containers affect how individuals move, and therefore can be inferred from the raw mobility data. 
Specifically, the amount of time spent within a container can depend on its hierarchical level. The connection between hierarchical level and mobility is supported by the literature, which shows that e.g.~transitions between regions are more frequent than transitions between countries\cite{amini2014impact}.

\textbf{A simple model identifies containers}.
We now describe the associated \textit{container model} of mobility, a model which estimates a person's containers from their empirical mobility patterns, see Figure~\ref{Figure1}c. For each individual, we model physical space as a hierarchy of $L$ levels, ordered from the smallest to largest (e.g.~individual locations to countries). At any level $l$, space is partitioned into topologically compact containers, with a characteristic size. For $l<L$, a container is fully included within a single parent container (for example each neighbourhood is part of a single city). Hence, each geographical location $k$ can be identified as a sequence of containers, $k  = ( k_1,..., k_l, ..., k_L)$, where container $k_{l}$ is included in $k_{l+1}$. 

Next, consistent with most models of human mobility\cite{barbosa2018human, fotheringham1983new}, each container $k_l$ is characterized by its probability to be selected within its parent container, its \emph{attractiveness} $a(k_l)$. 
We define the \emph{level-distance} $d(j,k)$ between locations $j$ and $k$ as the highest index at which the two sequences of containers describing $j$ and $k$ differ.\cite{saraccli2013comparison} We model traces individually; each trace results in a unique hierarchical structure.

Based on the assumption that the amount of time spent in a container depends upon its place in the hierarchy, we design a model of trajectories, where the probability of transitioning from location $j$ to location $k$ depends on the level-distance between them. 
For an agent located in $j$, we model the probability of moving to $k$ as the product of two factors: 
\begin{equation}
\label{equation1}
P( j \rightarrow k)= p_{d(j,k), d(j,h)} \prod _{l \leq d(j,k)}a(k_l),
\end{equation}
see also Methods, section `Model description'. 
The first factor, $p_{d(j,k), d(j,h)}$, represents the probability of traveling at level-distance $d(j,k)$, given that the current location $j$ is at level-distance $d(j,h)$ from the individual home-location, $h$. 
This probability follows a multinomial distribution, which must depend on level-distance from home to account for the fact that higher-level transitions are more likely when individuals are not in the home container; for example, one is typically more likely to transition at the country scale, when not in the home country.
The second factor $ \prod _{l \leq d(j,k)}a(k_l)$ is the probability of choosing a specific location $k$ at that level-distance, where $a(k_l)$ is the attractiveness of a container at level $l$ including location $k$.

\textbf{Results: Scales of human mobility}.
We fit this \textit{container model} to the individual GPS traces from two different datasets: dataset D1 which consists of traces of $\sim700\,000$ individuals distributed across the globe, and dataset D2 which consists of traces of $\sim 1000$ students from the Technical University of Denmark (see Methods section `Data description'). We fit the model using maximum likelihood estimation (see Methods section `Likelihood optimization'). 
For each individual, the fitting procedure outputs the most likely hierarchical spatial structure, along with attractiveness of containers and probabilities of traveling at given level-distance. 
We find that empirical individual mobility traces are characterized, on average, by four hierarchical levels. 
In contrast, synthetic traces generated by the current state-of-the-art, for example the \textit{Exploration and preferential return} (EPR) model \cite{song2010modelling} and its variations\cite{barbosa2015effect}, are best described by a single hierarchical level grouping individual stop locations (see Extended Data Figure~\ref{Extended_data_F5}). 
In both datasets of GPS traces, our model finds characteristic sizes of containers. The sizes of containers -- defined as the maximum distance between two locations in a container at a given level -- are not broad, but well described by lognormal distribution across the population. Our results are robust across datasets (see Extended Data Table~\ref{Extended_data_T1}). We argue that the characteristic sizes of containers are precisely the `scales' of human mobility.

These typical sizes of containers can be characterized by the median value $e^{\mu_l}$, of the lognormal distributions $\textrm{Lognormal}(\mu_l, \sigma_l^2)$\cite{gaddum1945lognormal}, for each hierarchical level $l$. We find $e^{\mu_2}=3.089 \pm 0.006 $~km, $e^{\mu_3}=27.064 \pm 0.006$~km, $e^{\mu_4}=88.442 \pm 0.022$~km and $e^{\mu_5}=161.634 \pm 0.049$~km, see Figure~\ref{Figure1}b and Extended Data Table~\ref{Extended_data_T3}. The coefficient of variation $C_l=\sqrt{e^{\sigma_l^2} - 1}$\cite{romeo2003broad}, characterizing the relative dispersion of the lognormal distribution, is included in the range [2.721,3.042] for $l$ in the range [2,5].

The median time spent within the same container at a given level is also well described by a lognormal distribution (see Figure~\ref{Figure1}c and Extended Data Table~\ref{Extended_data_T2}), implying that there are characteristic temporal scales associated to spatial scales. 

Having shown that we can infer information on geographical scales directly from the raw data, we now demonstrate the usefulness of this novel description of mobility patterns. We approach this task in two steps. First, we argue that the hierarchical description generated by the container model generalizes to unseen data without overfitting, while providing a more expressive and nuanced description of mobility relative to state-of-the-art models according to unbiased
performance estimates. Secondly, drawing on demographic and environmental data, we show that the container model produces results that converge with existing literature on gender differences, urban/rural divides, and walkability scores.

\begin{figure}[htb!]
\includegraphics[width=\textwidth]{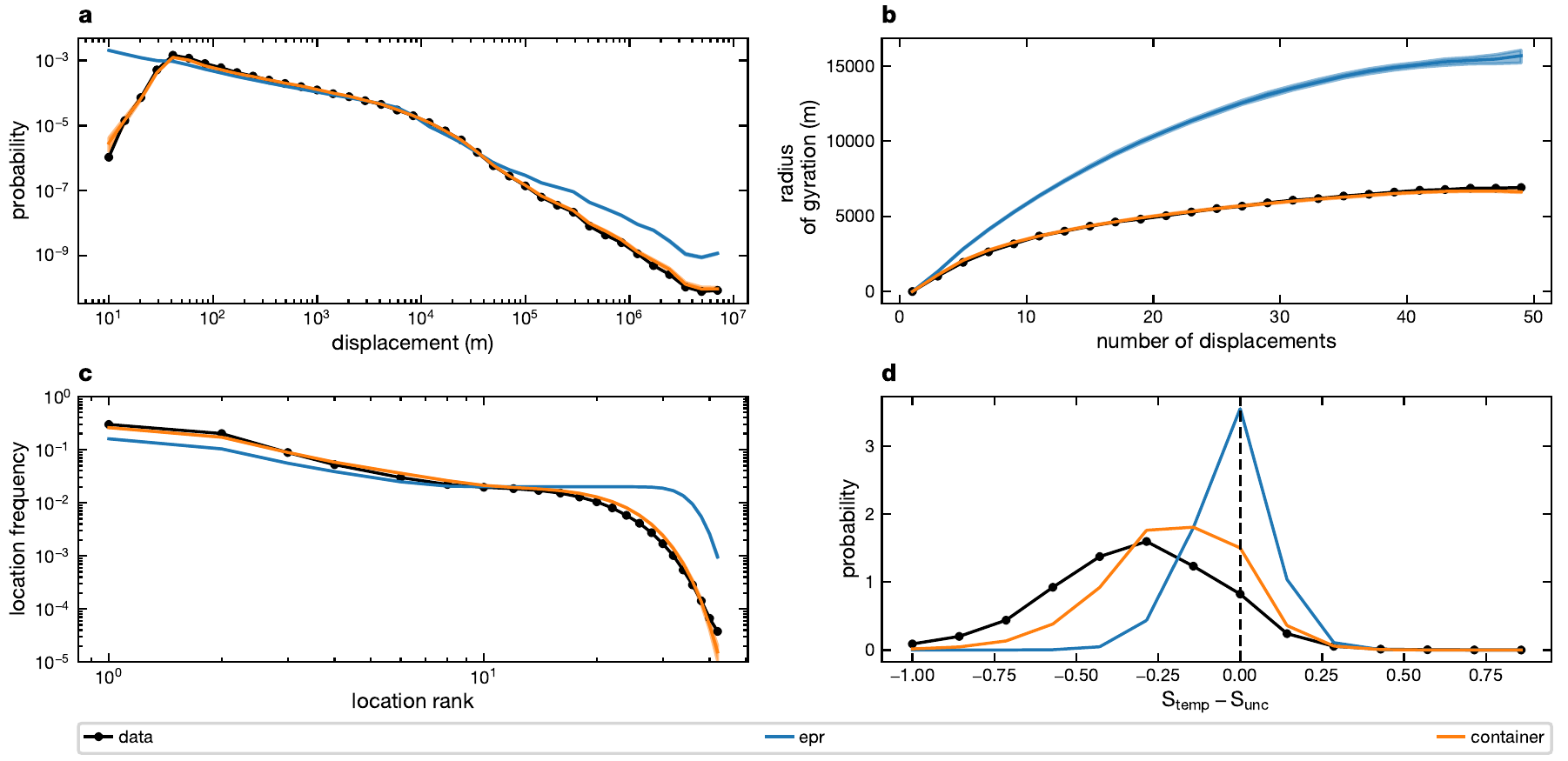}
\caption{\textbf{The container model generates realistic mobility traces.} (a) The distribution of displacements for the entire population, computed for real traces (black line, dots), traces generated by the container (orange filled area) and EPR\cite{song2010modelling} (blue filled area) models. (b) The median individual radius of gyration vs the number of displacements for data (black line, dots), traces generated by the container (orange filled area) and EPR (blue filled area) models. Dashed lines are logarithmic fits. (c) The average visitation frequency vs the rank of individuals' locations for real traces (black line, dots), container (orange filled area) and EPR (blue filled area) model traces. (d) The distribution of the difference between the temporal entropy $S_{temp}$ and the uncorrelated entropy $S_{unc}$ across individuals for real traces (black line, dots), and synthetic traces generated by the container (orange filled area) and the EPR model (blue filled area). In panels a,c,d, the filled areas for synthetic traces include two standard deviations around the mean computed across $1000$ simulations for each user. In panel b filled areas include the interquartile range. For each individual, we fitted the EPR and container models considering a training period of $1$ year. The data used here for validation corresponds to the $50$ individual displacements following the training period. Results are shown for a random sample of $9000$ individuals.}
\label{Figure2}
\end{figure}

\textbf{Validation: Generation of traces}. First, we explore the ability of the container model to capture key features of empirical mobility patterns and compare to state-of-the-art models. The container model allows us to generate synthetic traces. The realistic nature of these trajectories can be verified by comparing the statistical properties of synthetic and real sequences of locations (see Figure~\ref{Figure2}). For each individual, we fit the container model parameters using a portion of the entire trace with length $1$ year (see Methods section `Likelihood optimization'), and we then generate $1000$ synthetic sequences of $50$ displacements (see Methods section `Generation of traces'). 
Now, we can compare these synthetic traces with actual traces of the same length, collected in the $1$-year window subsequent to training. 
Thus, there is no overlap between the data we used to fit and to validate the model. Comparing synthetic traces to unseen data provides an unbiased performance
estimate, which allows us to compare model performance across multiple models
and confirm that the container model does not overfit (see Supplementary Note~4).

We focus on four key properties of mobility in the generated data: Distribution of displacements, evolution of radius of gyration, time allocation among locations, and entropy.

Considering the distribution of displacement lengths between consecutive locations, a widely studied property of mobility traces\cite{alessandretti2017multi}, the likelihood ratio test\cite{clauset2009power} shows that the container model provides a significantly better description of the data than the EPR model\cite{song2010modelling} and its variations (see Figure~\ref{Figure2}a and Extended Data Figure~\ref{Extended_data_F4} and Table~\ref{Extended_data_T4}, with $p\ll0.01$).

Next, the \textit{radius of gyration}\cite{gonzalez2008understanding} (see Methods section  `Metrics'), quantifies the spatial extent of an individual's mobility.
Here we find that while the evolution of individuals' radius of gyration over time is well described by a logarithmic growth in all cases: real\cite{gonzalez2008understanding}, EPR\cite{song2010modelling}, and the container model (see Figure~\ref{Figure2}b), only the fit $f(x) = a+b\cdot \log (x)$ for the container model traces is consistent with the real data within errors (see Supplementary Note~4).

We characterize the way in which individuals allocate time among locations (see Figure~\ref{Figure2}c), and find that distribution of location frequencies is better described by the container model, compared to the EPR model, under the likelihood ratio test\cite{clauset2009power} (with $p\ll0.01$). 

The final property of synthetic traces is the individual difference between the uncorrelated entropy $S_\textrm{unc}$, which characterizes the heterogeneity of visitation patterns, and the temporal entropy, $S_\textrm{temp}$, which depends not only on the frequency of visitation, but also the order in which locations were visited\cite{song2010limits} (see Methods section `Metrics'). 
The likelihood ratio test\cite{clauset2009power} shows that the distribution of $S_\textrm{unc}$ - $S_\textrm{temp}$ is better described by the container model, compared to the EPR model (with $p\ll0.01$). 
The result that the container model provides a better description of mobility compared to the state-of-the-art holds also when considering a comprehensive\cite{barbosa2018human} set of six state-of-the-art individual-level models (see Supplementary Note~4).

\begin{figure}[htb!]
\includegraphics[width=\textwidth]{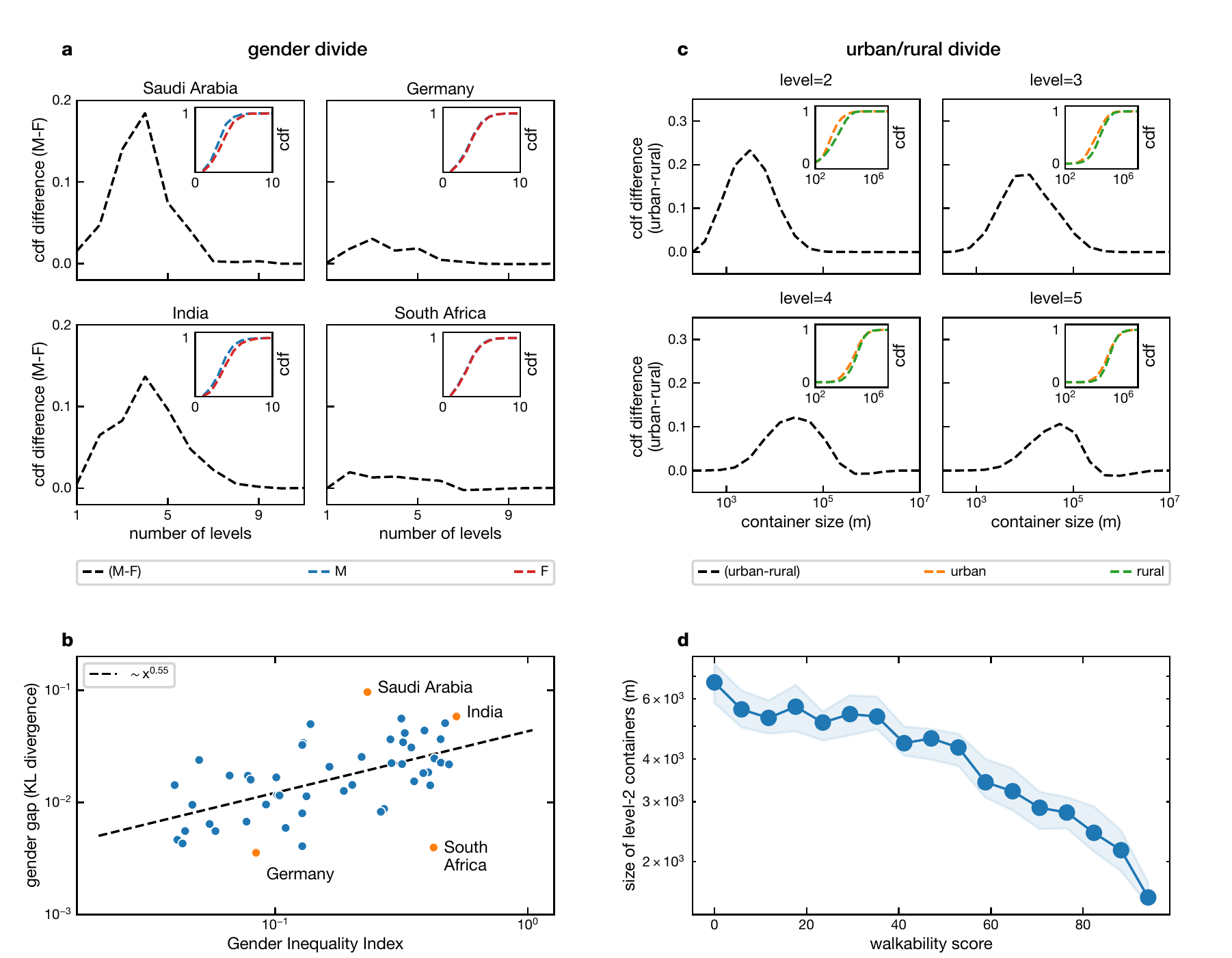}
\caption{\textbf{Socio-demographic differences and heterogeneity in scales.} (a) The cumulative distribution of number of levels for males (blue dashed line, inset) and females (red dashed line, inset), and the difference between the two (black dashed line). Results are shown for the four countries with the largest (left column, Saudi Arabia and India) and the smallest (right column, Germany and South Africa) gender gap, measured as the Kullback-Leibler (KL) divergence. (b) The gender gap in number of levels, computed as the  KL divergence between the number of levels for males and females, versus the Gender Inequality Index.\cite{gaye2010measuring} Each dot represents a different country, orange dots are the countries shown in (a). The black dashed line is a power law fit $P(x)\sim x^\beta$ with $\beta=0.58$. (c) The cumulative distribution of container sizes for individuals living in urban (orange dashed line, inset) and rural (green dashed line, inset) areas, and the difference between the two (black dashed line). Results are shown for hierarchical levels from $2$ to $5$. (d) The size of containers at level $2$ (with level $1$ corresponding to individual locations) versus the walkability score\cite{carr2010walk} around an individual's home location (blue dots). The shaded area corresponds to the $50\%$ interquartile range computed by bootstrapping 500 samples of individuals for each value of the walkability score.}
\label{Figure3}
\end{figure}

\textbf{Validation: Demographics and Built Environment}. Now, we aggregate users based on demographics and contextual features and explore the characteristics of containers for each subgroup of users, in order to underscore how the container model reveals patterns which have strong support in the existing literature. We focus on three factors which describe heterogeneity in mobility behavior: gender\cite{gauvin2020gender}, level of urbanization\cite{breheny1995compact}, and \textit{walkability} score\cite{carr2010walk} in the area surrounding one's home-location. 
First, we find that gender differences can partly explain the observed heterogeneity, in line with previous findings\cite{gauvin2020gender}, although not in all of the countries under study (see Figure~\ref{Figure3}a, and Supplementary Table~1). A novel finding concerns the fact that in $21$ out of $53$ countries, females are characterized by a significantly larger number of hierarchical levels than males, while the opposite is not the case for any country (see Supplementary Table~2). As a key observation inviting further research, we find that the difference between genders across countries, measured as the Kullback-Leibler divergence between the distributions of number of levels, is strongly correlated with the Gender Inequality Index (GII).\cite{gaye2010measuring} GII measures the percentage of potential human development loss due to gender inequality (Spearman correlation $\rho=0.69$, $p<10^{-6}$, see Figure~\ref{Figure3}b). Turning next to the size of containers, we find that in $48$ out of $53$ countries, the containers characterizing the mobility of females are smaller compared to males (see Supplementary Note~2), in line with previous research.\cite{gauvin2020gender}

Secondly, we find that the urban/rural divide partly explains differences in mobility patterns, in line with the fact that rural areas are characterized by limited accessibility.\cite{velaga2012transport, berry1967geography} Individuals living in rural areas (see Methods section `Other data' for definitions) have significantly larger containers compared to urban individuals, and this difference is more pronounced at the lowest hierarchical levels (see Figure~\ref{Figure3}c, and Supplementary Table~3).

Finally, we find that the walkability score around an individual's home location correlates negatively with the size of containers at the lowest hierarchical level (Spearman correlation $\rho=-0.96$, $p<10^{-9}$, see Figure~\ref{Figure3}d), in line with the finding that improved walkability increases accessibility to goods, services, and activities.\cite{litman2003economic} The correlation between walkability and container-size is significant up to the third level of description (see Supplementary Note~2).

\textbf{Discussion}.
The paradigm of power law descriptions does not stand entirely unchallenged within the quantitative analysis literature. For example, it has been argued that exponential or lognormal functions may be more suitable to describe the distributions of displacements within cities\cite{alessandretti2017multi}, hinting that human mobility may not be completely free of scales. 
Until now, however, the nature of the probability distribution of displacements has been unclear.\cite{alessandretti2017multi, gallotti2016stochastic}
For example, it has been suggested that scaling laws could be the signature of L\'{e}vy flights, a type of random walk with scale-free step-size attributed to animal foraging\cite{baronchelli2013levy}, but L\'{e}vy flights do not reproduce all statistical properties of human trajectories.\cite{song2010modelling}
It has also been proposed that the structure of the transportation system\cite{han2011origin, gallotti2016stochastic, zhao2015explaining}, where each mode of transportation corresponds to a typical distance traveled, could explain the observed scaling laws. 
Intra-city displacements considering all transportation modes, however, are not scale-free distributed, as this theory would suggest.\cite{liang2013unraveling, noulas2012tale}
Because of the lack of agreement on the functional form of distribution of displacements, many state-of-the-art agent-based models of individual mobility focus on temporal aspects\cite{barbosa2018human}, including the interplay between exploration and exploitation\cite{song2010modelling, pappalardo2015returners}, recency and memory effects\cite{alessandretti2018evidence, szell2012understanding, barbosa2015effect}, weekly and circadian rhythms.\cite{jiang2016timegeo}
With few exceptions\cite{han2011origin}, these models do not account for effects due to the spatial distribution of locations. 

In this paper, we have proposed a model in which human mobility is organized according to a hierarchical structure of spatial containers, corresponding to the notion of \textit{places} in geography (see Equation 1).
Under this model, the observed power law data arises by merging mobility \textit{within} containers with mobility that transports a person \textit{between} containers. The container model focuses on a specific aspect of mobility, and neglects other important features, including temporal visitation patterns, exploration\cite{song2010modelling}, and the structural connectedness of geographical spaces (e.g. through transportation networks)\cite{pumain2006alternative,batty2006hierarchy,arcaute2016cities}. These could be incorporated in future versions of the model. Fitting the model to trajectories collected in two distinct datasets, consisting of $\sim700\,000$ GPS traces of individuals distributed across the world, we found that -- across individuals -- the containers have typical sizes, representing the `scales' of human mobility.
We showed that our model allows for better understanding of mobility behavior and improves on the state-of-the-art in modeling.

\newpage
\section*{Methods}
\begin{small}
\subsection*{Data description and pre-processing}
\textbf{Mobility data. }Our analyses are based on two mobile phone datasets collecting high-resolution human trajectories. The study procedure follows the guidelines provided by the Danish Data Protection Agency.

The D1 dataset contains anonymized GPS location data for $\sim{5\,000\,000}$ individuals collected by a global smartphone and electronics company between $2017$ and $2019$ (see Extended Data Figure~\ref{Extended_data_F1}). The data consists of anonymized users who self-reported their age, gender, height, weight, and country of residence. Data was extracted through a smartphone app. All data analysis was carried out in accordance with the EU’s General Data Protection Regulation 2016/679 (GDPR) and the regulations set out by the Danish Data Protection Agency. We selected $\sim{700\,000}$ individuals with at least one year of data and whose position is known, every day, at least $50\%$ of the time. Individuals are located across the world and are aged between $18$ and $80$ years old, with an average age of $36$ years. About one-third of individuals are female. Gender and age were provided by the users at the time of registration. Data are not collected at a fixed sampling rate. Instead, the location estimate is updated when there is a change in the motion-state of the device (if the accelerometer registers a change). Location estimation error is below $100$~m for $93\%$ of data points. Informed consent was obtained for all study participants.

The D2 data was collected as part of an experiment that took place between September $2013$ and September $2015$.\cite{stopczynski2014measuring} The experiment involved $851$ Technical University of Denmark students ($\sim22\%$ female and $\sim78\%$ male), typically aged between $19$ and $21$ years old.  Participants’ position over time was estimated from a combination of GPS and WiFi information, resulting in samples every $1-2$ minutes. The location estimation error was below $50$~m in $95\%$ of the cases. Data collection was approved by the Danish Data Protection Agency. All participants provided informed consent. Data to produce Figure~\ref{Figure1}a is the location trajectory of one of the authors. We pre-processed all trajectories to obtain stop-locations using the Infostop algorithm.\cite{aslak2020infostop} We used the following algorithm parameters: \texttt{r1}$=30$~m, \texttt{r2}$=30$~m, \texttt{min\_staying\_time}$=10$ minutes, \texttt{max\_time\_between}$=24$ hours. Results are robust with respect to variation of these parameters (Supplementary Note~1).

\textbf{Other data. }We collected data on the walkability score in the area surrounding individuals' home locations using the WalkScore\cite{carr2010walk} API (\href{https://www.walkscore.com/professional/walk-score-apis.php}{https://www.walkscore.com/professional/walk-score-apis.php}). We collected data for $11,511$ individuals living in New Zealand, Australia, Canada and the USA, for which WalkScore data was available.

Data on the urbanization level in the area surrounding individuals' home locations is based on the GHS Settlement Model grid\cite{pesaresi2016operating} that delineates and classifies settlement typologies via a logic of population size, population and built-up area densities. This classification categorizes areas in urban areas, towns, and rural areas. In our analysis, we merged towns and cities into a single category. Data can be downloaded from: \href{https://ghsl.jrc.ec.europa.eu/data.php}{https://ghsl.jrc.ec.europa.eu/data.php}.

The Gender Inequality Index dataset can be downloded from:\\ \href{http://hdr.undp.org/en/content/gender-inequality-index-gii}{http://hdr.undp.org/en/content/gender-inequality-index-gii}. We used data for $2017$.

\subsection*{The container model}

\textbf{Model description. } The container model models the trace of an agent transitioning between locations in space. The model is specified by three sets of parameters that can either be simulated to generate synthetic traces or estimated for an empirical trace through maximum likelihood estimation. The model contains:
\begin{itemize}
\item A hierarchical structure $\boldsymbol{H}$ with $L$ levels, where each level consists of containers encapsulating locations. Accordingly, each location $k$ can be described as a sequence of containers encapsulated within each other, $k  = ( k_1,..., k_l, ..., k_L)$, where levels are ordered from the most fine grained $l=1$ to the most coarse grained $l=L$. In analogy, a restaurant can be described as a sequence corresponding to the building, the neighbourhood, the city, etc. where it is located. At each level in the hierarchy, containers have comparable size. In the simplest form, this structure is a nested grid (see Supplementary Note~3).
\item The collection of these containers' attractivenesses, $\boldsymbol{a}$. The attractiveness $a(k_l)$ is the probability of visiting container $k_l$ among all containers encapsulated within $k_{l+1}$. Accordingly, $\sum_{k_l \in k_{l+1}}a(k_l)=1$. 
\item The $L \times L$ matrix, $\boldsymbol{p}$ characterizing the probability of travelling at a certain level-distance. Each row in $\boldsymbol{p}$ is a probability vector that describes the probabilities $p(d,d_h)$ of traveling at level-distance $d$ when the level distance from home is $d_h$. Here, home is defined as the location with largest attractiveness at all levels. By level-distance we mean the so-called cophenetic distance\cite{saraccli2013comparison}: the highest level in the hierarchy one travels to reach the destination. It is necessary to maintain separate probability distributions for each level-distance from home. This is, for example, because traveling at the highest-level distance (e.g. intercontinentally) is unlikely when one is near home, but comparatively likely when on a different continent.
\end{itemize}
Under the container model, each transition is the result of a two-stage decision process. 
First, the individual selects at which level-distance to travel. 
Then, she selects a specific destination based on container attractiveness. Specifically, an individual located in $j$, chooses destination location $k$ with probability:
\[
 P_{\boldsymbol{H}, \boldsymbol{a}, \boldsymbol{p}}( j \rightarrow k )= p_{d(j,k), d(j,h)} \frac{a(k_{d(j,k)})}{1 - a(j_{d(j,k)})} \prod_{l=1}^{d(j,k)-1} a(k_l)
\]
The first factor, $p_{d(j,k), d(j,h)}$, is the probability of traveling at level-distance $d(j,k)$. 
The second factor $\frac{a(k_{d(j,k)})}{1 - a(j_{d(j,k)})}$ is the probability of choosing container $k_{d(j,k)}$. 
Such a container is found at level $d(j,k)$ in the hierarchy and has attractiveness $a(k_{d(j,k)})$. 
The renormalization $1-a(j_{d(j, k)})$ accounts for the fact that container $j_{d(j, k)}$ cannot be selected (this detail is not present in the main text for readability).
The third factor $\prod_{l=1}^{d(j,k)-1} a(k_l)$ is the probability of picking all other containers $k_l$ that encapsulate location $k$, for any level in the hierarchy lower than $d(j,k)$. Note that the way we model destination choice in a hierarchical fashion, connects to the class of choice models called nested logit models.\cite{train2009discrete} The nested structure of the physical space in the container model relates, in part, to the organization of the transportation system\cite{marchetti1994anthropological,zahavi1978stability,miller2004tobler} and to the concrete structure of our built environment\cite{berry1967geography,batty2006hierarchy}. The importance of these contexts are also gradually being recognized in the Human Mobility literature, where early studies focused on large datasets, but did not consider the effect of contextual information, e.g. transportation type or other mobility characteristics, which can introduce heterogeneity.\cite{noulas2012tale, zhao2015explaining, gauvin2020gender, kraemer2020mapping, steele2017mapping, lu2016detecting, lu2012predictability, weiss2018global, althoff2017large}

\textbf*{Generating traces. }We model transitions as a two-step decision process. Thus, we can simulate synthetic trajectories given a hierarchical description $\boldsymbol{H}$, container attractivenesses $\boldsymbol{a}$, and the probability matrix $\boldsymbol{p}$, (either designed or obtained by fitting the container model to an empirical trace). 
We simulate the mobility of an agent by the following algorithm. To guide the reader we offer an \emph{Example} at each step, describing an agent travelling across a hierarchy where levels correspond to countries, cities, neighbourhoods, buildings and locations.
\begin{enumerate}
    \item Initialize the agent in a random location, $j$, at level-distance $d(j, h)$ from the \textit{home} location. \emph{Example: the agent is initialized in location $j$ located in a different country than her home country.}
    
    \item Select a level-distance $l^*$ that the agent should travel at, by drawing from the multinomial distribution, $p_{d(j,h)}$. \emph{Example: the agent chooses to travel at the city-distance.}
    
    \item Select a destination, $k$: 
    \begin{itemize}
    \item {At level $l^*$, select a container $k_{l^*}$, by drawing from the attractiveness distribution over the containers encapsulated in $j_{l^*+1}$. $j_{l^*}$ cannot be selected in this process, so $k_{l^*} \neq j_{l^*}$. \emph{Example: the agent chooses the destination city among other cities in the same country where she is currently located.}}
    \item {At level $(l^*-1)$, select a container $k_{l^*-1}$ encapsulated within $k_{l^*}$, by drawing from the attractiveness distribution over containers in $k_{l^*}$. Continue this process until level 1 is reached. \emph{Example: the agent picks a neighborhood, then a building and then a location encapsulated within the destination city chosen in the previous step.}}
    \end{itemize}
    \item Repeat steps 2 and 3 for any desired number of displacements.
\end{enumerate}

\textbf{Likelihood optimization.} We can fit the container model to an empirical trace and obtain the model parameters $\boldsymbol{H}$, $\boldsymbol{a}$, $\boldsymbol{p}$, using maximum likelihood estimation. We write the likelihood that a sequence of individual locations $T=[k(0),...,k(i),...k(n_T)]$ was generated by an instance of the container model specified by $\boldsymbol{H}$, $\boldsymbol{a}$, $\boldsymbol{p}$, as: 
\[
\pazocal{L}(\boldsymbol{H}, \boldsymbol{a}, \boldsymbol{p} \mid  T ) = \prod_{i=0}^{n_T-1} P_{\boldsymbol{H}, \boldsymbol{a}, \boldsymbol{p}}(k(i-1) \rightarrow k(i)),
\]
where $P_{\boldsymbol{H}, \boldsymbol{a}, \boldsymbol{p}}(k(i-1) \rightarrow k(i))$ is the probability of a transition to occur. 

Unlike spatial clustering methodologies, this method allows us to identify `containers', structures that are compact in size, but also also contain mobility behavior. 
This optimization task, however, is computationally expensive, therefore we approach the problem according to the following heuristic.

First, we note that, when $n_T$ is large and $\boldsymbol{H}$ is chosen, $\boldsymbol{p}$ and $\boldsymbol{a}$ are trivial to estimate. In fact, for $n_T \to \infty$, element $p_{d,d_h}$ of matrix $\boldsymbol{p}$ equals the fraction of transitions covering a level-distance $d$ among all transitions starting at level-distance $d_h$ from home. 
Similarly, for $n_T \to \infty$, the attractiveness of a container equals the fraction of times such container is selected among containers in the same parent container. 

Thus, for long enough traces, we can estimate the maximum likelihood by exploring different choices of $\boldsymbol{H}$ only, where $\boldsymbol{H}$ is effectively a spatial hierarchical partition of individual locations. 
In order to ensure that clusters are compact, we choose $\boldsymbol{H}$ among the solutions of the complete linkage hierarchical clustering algorithm.\cite{everitt2011hierarchical}

First, we run the complete-linkage algorithm for the set of locations in sequence $T$. 
The algorithm initializes each location as a separate cluster. 
It then iteratively joins the two clusters whose union has the smallest diameter, defined as the maximum distance between two locations in a cluster, and stores the clustering solution. 
It runs for $N$ iterations, where $N$ is the number of locations (and possible clustering solutions). In the final iteration all locations form one cluster. 
The result of the complete-linkage algorithm can be visualized as a dendogram and queried for clusters at any cut-distance (see Extended Data Figure~\ref{Extended_data_F3}a).

We then proceed to find the hierarchical partition $\boldsymbol{H^*}$ corresponding to the maximum likelihood $\pazocal{L}^*$. 
Exhaustive search would require computing the likelihood for all possible partitions $\boldsymbol{H}$.
When we let $L$ range from $1$ to $N$, we arrive at the total number of possible partitions by the following logic: 
For $L=1$, the dendogram is cut zero times so there is one partition, which has only individual locations and no containers; for $L=2$ the dendogram is cut once, so there are $N$ partitions because there are $N$ ways to cut the dendogram; for $L=3$, there are $\frac{N(N-1)}{2}$ ways to cut the dendogram two times, and so on. The set of all possible partitions then has size $\sum_{L=1}^{N} {\binom{N}{L}}$. 

We define a heuristic to reduce the set of candidate partitions $\boldsymbol{H}$, by optimizing the likelihood \emph{one-level-at-a-time}, see Extended Data Figure~\ref{Extended_data_F3}b. 
The algorithm works as follows: 
First, we compute the likelihood $\pazocal{L}_{1}$ of $T$ in the case $L=1$, corresponding to having no containers. 
Then, we test the $N$ possible partitions corresponding to cutting the dendogram one time (i.e. $L=2$), by computing the corresponding likelihoods. 
We find the cut $C_2$ of the dendogram resulting in the maximum likelihood $\pazocal{L}_2$. If $\pazocal{L}_{1}$ is significantly larger than $\pazocal{L}_{2}$ (tested by bootstrapping), we assign $\pazocal{L}^*=\pazocal{L}_{1}$, conclude that $\boldsymbol{H^*}$ has only $1$ level (individual locations), and stop the algorithm. 
Otherwise, we explore the set of partitions corresponding to two cuts of the dendogram (i.e. $L=3$), where one of them is $C_2$, and find the cut $C_3$ that yields the maximum likelihood $\pazocal{L}_3$. 
We compare $\pazocal{L}_2$ and $\pazocal{L}_3$, and stop the algorithm if $\pazocal{L}_2$ is significantly larger than $\pazocal{L}_3$. 
We proceed for increasing values of $L$, until $L=N$ or there is no significant improvement in likelihood. 
In the worst-case scenario, we explore $N!$ partitions.
We validate the \emph{one-level-at-a-time} algorithm against synthetic data (see Extended Data Figure~\ref{Extended_data_F3} and Supplementary Note~3). We find that the algorithm recovers the correct number of hierarchical levels $\sim95\%$ of the times. 
The similarity between the correct and recovered hierarchical structure, measured as their cophenetic correlation\cite{saraccli2013comparison} has median value $1$ (the cophenetic correlation is the correlation between the cophenetic distance computed for all pairs of locations according to two different hierarchical descriptions, and thus is $1$ for identical descriptions). The median absolute error relative to the estimation of the matrix of probabilities $\boldsymbol{p}$ is $0.03$.

\subsection*{Model validation}

\textbf{Metrics.} In Figure~\ref{Figure2}, we compare synthetic and real traces by computing quantities characterizing individual trajectories. \\

The radius of gyration for an individual $u$ is defined as:
\[
r_g^{u} = \sqrt{\frac{1}{N} \sum_{n=0}^{N} (r_n^u - r_{CM,n}^u)^2},
\] 
where $N$ is the total number of displacements ($50$ in our analysis), $r_n^u$ is the position of $u$ after $n$ displacements, $r_{CM, n}^u$ is its center of mass after $n$ displacements, defined as:
\[
r_{CM,n}^u = \frac{1}{n}\sum_{j=0}^{n} r_j^u.
\]
The uncorrelated entropy $S_{unc}$ is defined as:
\[
S_{unc} =  - \sum_{i=0}^{N_L} P(i) log_2(P(i)),
\]
where $P(i)$ is the probability of visiting location $i$, and $N_L$ is the total number of locations. The temporal entropy $S_{temp}$, is defined as 
\[
S_{temp} =  - \sum_{T_{i'} \in T_{i}} P(T_{i'}) log_2(P(T_{i'})),
\] 
where $P(T_{i'})$ is the probability of finding a particular time-ordered subsequence $T_{i'}$ in the trajectory $T_i$. We estimate $S_{temp}$ using the method described by Sekara \textit{et al}.\cite{sekara2016fundamental}

\textbf{EPR model. }We generate EPR synthetic traces as follows. 
First, we fit the model parameters~\cite{song2014prediction} and determine, for each individual, the number of visited locations $S$ as well as the number of visits $f_i$ per location $i$ using traces with one-year duration. Then, we generate traces using the model described in Song \textit{et al}.\cite{song2014prediction}
At each new displacement, an individual explores a new place with probability $\rho S^{-\gamma}$ and exploits a previously known location with the complementary probability. 
In the first case, she chooses a place at distance $\Delta r$, extracted from a power law distribution $P(\Delta r)\sim \Delta r ^{-\beta}$. 
In the latter case, she chooses a previously visited location $i$ with probability proportional to the number of visits $f_i$. See Supplementary Note~4 for further details and the implementation of other models.

\subsection*{Declarations}
\textbf{Data Availability.} Derived data that support the findings of this study are available in DTU Data with the identifier \href{https://doi.org/10.11583/DTU.12941993.v1}{https://doi.org/10.11583/DTU.12941993.v1}. Source data for Figures 1, 2 and 3 are provided with the paper. Additional data related to this paper may be requested from the authors. Raw data for dataset D1 are not publicly available to preserve individuals' privacy under the European General Data Protection Regulation. Raw data for dataset D2 are not publicly available due to privacy considerations, but are available to researchers who meet the criteria for access to confidential data, sign a confidentiality agreement, and agree to work under supervision in Copenhagen. 
Please direct your queries to the corresponding author.

\textbf{Code Availability.} \\
Code is available at \href{https://github.com/lalessan/scales\_human\_mobility/}{https://github.com/lalessan/scales\_human\_mobility/}\\
An interactive illustration of the \emph{container model} generative process can be found at \\ \href{https://observablehq.com/@ulfaslak/a-model-for-generating-multiscale-mobility-traces}{https://observablehq.com/@ulfaslak/a-model-for-generating-multiscale-mobility-traces}.
For a visual demonstration of how scale-free distribution can emerge from the aggregation of other distributions, refer to \href{https://observablehq.com/@ulfaslak/a-visual-exploration-of-how-a-power-law-can-emerge-from-aggre}{https://observablehq.com/@ulfaslak/a-visual-exploration-of-how-a-power-law-can-emerge-from-aggre}.

\textbf{Acknowledgements.} We thank Filippo Simini and Lars Kai Hansen for providing insightful comments. We thank Marta C. Gonzalez for help with datasets. 

\textbf{Author Contributions.} LA, UA and SL designed the study and the model. LA and UA performed the analyses and implemented the model. LA, UA and SL analysed the results and wrote the paper. 

\textbf{Corresponding Author.} Correspondence and requests for materials should be addressed to Sune Lehmann~(email: sljo@dtu.dk).

\end{small}

\bibliographystyle{naturemag}
\bibliography{biblio_scales}

\clearpage
\renewcommand{\thetable}{S\arabic{table}}%
\renewcommand{\thefigure}{S\arabic{figure}}%
\renewcommand{\thesubsection}{Supplementary Note \arabic{subsection}:}%
\section*{Supplementary Notes}
\subsection{Containers have typical scales.}\label{Note1}

\textbf{Definition of scale-free distribution. } Power-law distributions are called `scale-free' (or scale invariant) distributions because they look the `same' regardless of the scale at which they are looked at. 
Formally, we define a distribution function $f$ as `scale-free' if there exists an $x_0>0$ and a continuous function $g$ such that $f(\lambda x) = g(\lambda)f(x)$ for all $x,\lambda$, satisfying $\lambda x \geq x_0$.\\
This definition suggests that the shape of the distribution $f$ remains unchanged, up to a multiplicative factor $g(\lambda)$, when $x$ is scaled by $\lambda$. 
It can be proven that a distribution function $f$ satisfies the definition above if and only if it has a power law tail, i.e., there exists $x_0>0, c \geq 0$, and $\alpha \geq 0$ such that $f(x) = c x^{-\alpha}$ for $x \geq x_0$.\cite{newman2005power} \\
For a visual demonstration of how scale-free distribution can emerge from the aggregation of other distributions, refer to \href{https://observablehq.com/@ulfaslak/a-visual-exploration-of-how-a-power-law-can-emerge-from-aggre}{https://observablehq.com/@ulfaslak/a-visual-exploration-of-how-a-power-law-can-emerge-from-aggre}.

\textbf{Empirical distributions of containers are not scale-free.} The sizes of containers characterizing human data -- defined as the maximum distance between two locations in the same container at a given level -- are not broad, but better described by lognormal distribution across the population under study. This is verified by considering a truncated power-law distribution:
\[
f_1(x,a,\lambda) = \frac{\lambda^{a+1} x^{a} e^{-\lambda x}}{ \Gamma(a)}
\]
with parameters $a$ and $\lambda$, and where $\Gamma$ is the gamma function $\Gamma(x) = (x-1)!$, in comparison with a lognormal distribution:

\[
f_2(x,\sigma,\mu) = \frac{1}{xs\sqrt{2\pi}}e^{-\frac{(\log x- \mu)^2}{2\sigma^2}}
\]

with parameters $\sigma$ and $\mu$. We compare the goodness-of-fit of distributions $f_1$ and $f_2$ using the log-likelihood ratio test.\cite{clauset2009power} We find that the lognormal distribution are better suitable, compared to truncated power laws, to describe the distribution of container sizes at levels $2$ to $5$ at significance level $\alpha=0.01$ (see Extended Data Table~1 and Extended Data Figure~2), for both datasets. The same result is verified for the distribution of time spent within the same container at a given level (see Extended Data Table~2). Further, we find that the log-normal distribution is generally a better model of the containers size distribution in comparison to other log-well-behaved distributions: the log-gamma, the log-logistic, the log-laplace and the log-Weibull distributions (see Extended Data Table~1). In Extended Data Table~3, we report the characteristics of the fitted lognormal distributions. Note that our results are supported by the empirical finding that the distribution of population clusters (corresponding to densely populated areas) is lognormal across a large range of values.\cite{eeckhout2004gibrat,corral2020truncated}

\textbf{Results are robust across different definitions of stop-location.}
Our smallest spatial unit of analysis is the stop-location. We verified that the distributions of container sizes and time spent within container are best described by a log-normal distribution for different definitions of stop-location. In particular, we compared the results obtained after running the stop-location detection algorithm\cite{aslak2020infostop} for different values of the parameters $r1$ and $r2$ characterizing the size of stop-locations. Results obtained for dataset $D2$ and values of $r1$ and $r2$ in  [$10$, $20$, $30$, $50$] (meters) are presented in Extended Data Table~1.

\subsection{Demographics and Built Environment}
\textbf{Gender differences revealed by the container model.\label{Section5}}
We investigate gender differences with respect to mobility features revealed by the container model, by considering countries in our dataset that include more than $1000$ individuals. First, we investigate the divergence between the number of hierarchical levels for males vs females. We find that the gender divide, measured as the Kullback-Leibler divergence between the distributions, is considerably more pronounced in certain countries compared to others, ranging from $0.09$ in Saudi Arabia to $0.0036$ in Germany (see Supplementary Table~\ref{TableGender}). The gender divide in number of levels correlates with the gender inequality index (see main text, Figure~2B). In fact, under the Kolmogorov–Smirnov test, only in $21$ countries out of $53$ the difference is significant. We find that in $44$ countries, females mobility has more hierarchical levels than males', under the Mann–Whitney U test (see Supplementary Table~\ref{TableGender}). The opposite is not verified in any country (see Supplementary Table~\ref{TableGender}). Then, we study differences in containers sizes. We find that in $48$ countries out of $53$, the size of level-$2$ containers is smaller for females compared to males, under the Mann–Whitney U test (see Supplementary Table~\ref{TableGender2}). The Kullback-Leibler divergence between females and males distributions of level-$2$ containers sizes correlates positively with the Gender Inequality Index ($\rho =0.40$, $p<10^{-2}$).

\textbf{Urban/rural differences revealed by the container model.\label{Section6}}
We investigate differences between users living in urban vs rural areas with respect to the size of containers. Data on the urbanisation level in the area surrounding individuals' home locations relies on the GHS Settlement Model grid. For the sake of our analysis, we merged towns and cities into the same category (urban). We find that, at all hierarchical levels, the size of containers is significantly larger for individuals living in rural areas compared to urban individuals (see Supplementary Table~\ref{TableUrban}) under the Mann–Whitney U test. Differences between the distribution of container sizes, measured as Kullback-Leibler divergence, are more pronounced at low hierarchical levels (see Supplementary Table~\ref{TableUrban}).

\textbf{Differences reflecting the walkability around one's home location. \label{Section7}}
We investigate if the walkability score in the area surrounding an individual's home location affects the size of containers revealed by our model. We find that there is a negative correlation between the size of containers and the walkability score, for containers at level $2$ ($\rho=-0.98$, $p<10^{-9}$, see main text, Figure~2D) and $3$ (Spearman correlation $\rho=0.68$, $p=0.001$). The correlation is not significant at significance level $\alpha=0.01$ for higher hierarchical levels. These analyses are computed for the $11,511$ individuals living in New Zealand, Australia, Canada and the USA, for which WalkScore data was available (see Material and Methods, Data Description).

\subsection{Evaluation of the likelihood optimization algorithm. \label{sec:simulations}}
In this section, we evaluate the performance of the likelihood maximization algorithm for the container model (see main text, Methods, Model description). First we generate $5,000$ synthetic traces using the container model as follows:
\begin{itemize}
    \item We define a hierarchical partitioning $\boldsymbol{H}$ as a nested grid with $L$ levels, where at each level $i$ there are $N_i$ squared containers. $L$ and all values of $N_i$ are random integers (from a uniform distribution). Each cell in the grid corresponds to a location. 
    \item We fill the matrix $\boldsymbol{p}$ with values chosen from a uniform distribution, and we then normalize it appropriately (to ensure that values correspond to probabilities of travelling at given level-distance). Note that here, for simplicity, we consider that the probability of travelling at given level-distance does not depend on the distance from home.
    \item We assign to containers their attractiveness (stored in $\boldsymbol{a}$) as follows. The distribution of attractiveness for the children of any given parent container is a power law (with exponent chosen from a uniform distribution), in agreement with the literature.\cite{barbosa2018human}
    \item Given the above choices of $\boldsymbol{H}$, $\boldsymbol{a}$ and $\boldsymbol{p}$, we simulate sequences of $2,000$ locations following the procedure described in the main text (Methods, Generation of traces).
\end{itemize}

An interactive illustration of this generative process can be found at \\ \href{https://observablehq.com/@ulfaslak/a-model-for-generating-multiscale-mobility-traces}{https://observablehq.com/@ulfaslak/a-model-for-generating-multiscale-mobility-traces}. We then estimate $\boldsymbol{H}$, $\boldsymbol{a}$ and $\boldsymbol{p}$ using the \emph{one-level-at-a-time} algorithm for maximum likelihood optimization described in the main text (Methods, Likelihood optimization). We evaluate the performance of the algorithm by computing: the absolute difference $D$ between the original and recovered number of levels (we find that P($D=0$)=0.95, P($D\leq 1$)=0.99), the cophenetic correlation\cite{saraccli2013comparison} between the original and recovered hierarchical structure (we find that the cophenetic correlation has median value $1$), and the relative difference $\frac{|p_{o} - p_{r}|}{p_r}$ between the original $p_{o}$ and recovered $p_{r}$ values in matrix $\boldsymbol{p}$ (that has median value $0.03$). Results are shown in Extended Data Figure~3.

\subsection{Comparison across models. \label{sec:comparison_models}}

We compare the ability to reproduce the statistical properties of human mobility of several individual-level models: the container model presented here, the Exploration and Preferential Return model (epr)\cite{song2010modelling}, the memory EPR (m-epr)\cite{alessandretti2018evidence}, the gravity EPR (d-epr)\cite{pappalardo2016human}, the DITRAS model\cite{pappalardo2018data}, and the recency EPR~(recency-epr).\cite{barbosa2015effect}

\textbf{Implementation. }For each individual, we consider 1 year of data to train the model, and 50 displacements to test the performance of the model. 
We estimate the parameters by fitting the training data. For all models but DITRAS, we trained each individual model by using the training data to find the model parameters and build the visitation frequency of locations (the implementation of the models can be found at\\ \href{https://github.com/lalessan/scales\_human\_mobility/}{https://github.com/lalessan/scales\_human\_mobility/}). We used the implementation of DITRAS of scikit-mobiliy.\cite{pappalardo2019scikit} Note that the Ditras and d-epr model are not implemented for dataset D2, because the number of unique locations in the entire dataset is extremely large.

\textbf{Evaluation} We consider four properties of human mobility: the probability of displacement lengths $(P\Delta r)$, the distribution of number of visits to each location $P(visits)$, the distribution of the difference between the temporal and uncorrelated entropy $P(S_{unc} - S_{temp})$, and the temporal evolution of the radius of gyration. For each individual, we measure the likelihood of the data in the test set given the probability distribution observed in the synthetic data (see Extended Data Table~4). Using the log-likelihood ratio test\cite{clauset2009power}, we find that the container model provides a better description than the others across the measures considered at significance level $\alpha=0.01$. Further, we find that while the evolution of individuals’ average radius of gyration over time is well described by a logarithmic growth in all cases, only the fit f(x) =a+b·log(x) for the container model traces is consistent with the real data within errors: ($a_\textrm{data}=-628\pm 138$, $b_\textrm{data}=1884 \pm44$, $a_\textrm{container} = -523 \pm 27$, $b_\textrm{container} = 1844 \pm 9$), while the EPR model fit has markedly different parameters ($a_\textrm{EPR}=-5700 \pm 92$, $b_\textrm{EPR}=6087 \pm 36$).

Finally, fitting the container model to synthetic traces, we find that synthetic traces generated by the epr model are characterized, on average, by a single hierarchical level encapsulating individual locations (see Extended Data Figure~5). Instead empirical individual mobility traces are characterized, on average, by four hierarchical levels (see Extended Data Figure~5).

\newpage
\section*{Supplementary Tables}
\begin{table}[h!]
\begin{scriptsize}
\begin{minipage}[t]{0.5\linewidth}
\begin{tabular}[t]{llrlll}
\toprule
{} &               country & KL & KS test & MW test  & MW test  \\
{} &                &  &                &              (M$>$F)&   (F$>$M)\\
\midrule
1  &          Saudi Arabia &        0.0962 &                    **** &                        ns &                      **** \\
2  &                 India &        0.0583 &                    **** &                        ns &                      **** \\
3  &               Romania &        0.0559 &                    **** &                        ns &                      **** \\
4  &                Jordan &        0.0509 &                       * &                        ns &                        ** \\
5  &                Emirates &        0.0498 &                      ns &                        ns &                         * \\
6  &               Ecuador &        0.0437 &                      ** &                        ns &                       *** \\
7  &           South Korea &        0.0422 &                       * &                        ns &                       *** \\
8  &                 Chile &        0.0415 &                      ns &                        ns &                         * \\
9  &                 Egypt &        0.0366 &                      ** &                        ns &                      **** \\
10 &              Malaysia &        0.0365 &                      ns &                        ns &                        ** \\
11 &                Turkey &        0.0344 &                      ns &                        ns &                      **** \\
12 &             Hong Kong &        0.0340 &                    **** &                        ns &                      **** \\
13 &             Lithuania &        0.0339 &                      ** &                        ns &                      **** \\
14 &               Croatia &        0.0325 &                      ns &                        ns &                       *** \\
15 &                Mexico &        0.0309 &                    **** &                        ns &                      **** \\
16 &              Bulgaria &        0.0254 &                      ns &                        ns &                         * \\
17 &           Philippines &        0.0246 &                      ns &                        ns &                         * \\
18 &                Taiwan &        0.0241 &                    **** &                        ns &                      **** \\
19 &               Belgium &        0.0239 &                      ns &                        ns &                        ns \\
20 &             Indonesia &        0.0226 &                       * &                        ns &                        ** \\
21 &               Ukraine &        0.0224 &                      ns &                        ns &                        ** \\
22 &               Vietnam &        0.0220 &                      ns &                        ns &                        ** \\
23 &                  Iran &        0.0218 &                      ns &                        ns &                        ** \\
24 &                 China &        0.0208 &                      ns &                        ns &                        ns \\
25 &                Brazil &        0.0185 &                    **** &                        ns &                      **** \\
26 &              Thailand &        0.0182 &                      ns &                        ns &                       *** \\
 27 &             Singapore &        0.0174 &                      ns &                        ns &                        ** \\
\end{tabular}
\end{minipage}
\begin{minipage}[t]{0.5\linewidth}
\begin{tabular}[t]{|llrlll}
\toprule
{} &               country & KL & KS test & MW test  & MW test  \\
{} &                &  &                &              (M$>$F)&   (F$>$M)\\
\midrule

28 &               Austria &        0.0173 &                       * &                        ns &                        ** \\
29 &               Ireland &        0.0167 &                      ns &                        ns &                        ns \\
30 &                 Italy &        0.0160 &                       * &                        ns &                        ** \\
31 &             Argentina &        0.0154 &                      ns &                        ns &                         * \\
32 &         United States &        0.0143 &                      ** &                        ns &                      **** \\
33 &           Switzerland &        0.0143 &                      ns &                        ns &                        ns \\
34 &              Colombia &        0.0142 &                      ns &                        ns &                         * \\
35 &              Slovakia &        0.0127 &                      ns &                        ns &                        ** \\
36 &                 Japan &        0.0116 &                    **** &                        ns &                      **** \\
37 &             Australia &        0.0115 &                      ns &                        ns &                        ns \\
38 &        Czech Republic &        0.0113 &                       * &                        ns &                       *** \\
39 &                Canada &        0.0096 &                      ns &                        ns &                        ns \\
40 &                Norway &        0.0095 &                      ns &                        ns &                         * \\
41 &               Hungary &        0.0088 &                      ns &                        ns &                         * \\
42 &                Russia &        0.0083 &                      ns &                        ns &                      **** \\
43 &                Poland &        0.0080 &                      ns &                        ns &                       *** \\
44 &                 Spain &        0.0067 &                      ns &                        ns &                        ** \\
45 &               Finland &        0.0064 &                      ns &                        ns &                         * \\
46 &               Estonia &        0.0059 &                      ns &                        ns &                        ns \\
47 &           Netherlands &        0.0056 &                      ns &                        ns &                       *** \\
48 &                France &        0.0055 &                     *** &                        ns &                      **** \\
49 &               Denmark &        0.0046 &                      ns &                        ns &                        ns \\
50 &                Sweden &        0.0043 &                      ns &                        ns &                      **** \\
51 &        United Kingdom &        0.0041 &                    **** &                        ns &                      **** \\
52 &          South Africa &        0.0040 &                      ns &                        ns &                        ns \\
53 &               Germany &        0.0036 &                    **** &                        ns &                      **** \\
 &               &         &                       &                        &                       *** \\
\end{tabular}
\end{minipage}
\end{scriptsize}
\caption{\textbf{Gender differences in number of levels.} Statistics concerning the difference between the distribution of the number of levels for males and females users in countries with more than $1000$ users. We report the Kullback-Leibler divergence (KL), as well as the p-values associated to the Kolmogorov–Smirnov test (KS test), and the Mann–Whitney U test testing that a randomly selected value from the male population is greater than a randomly selected value from the female population (MW test, M$>$F) and vice-versa (MW test, F$>$M). We used the following notation for p-values:  p$>$ 0.05 (ns),  p$\leq$ 0.05 (*), p$\leq$ 0.01 (**), p$\leq$ 0.001 (***), p$\leq$ 0.0001 (****)\label{TableGender}}
\end{table}

\clearpage

\begin{table}
\begin{scriptsize}

\begin{tabular}{llrlll}
\toprule
{} &               country & KL & KS test & MW test & MW test  \\
{} &                &  &              & (M$>$F)   &                   (F$>$M)  \\
\midrule
1  &          Saudi Arabia &        0.2121 &                    **** &                      **** &                        ns \\
2  &                Jordan &        0.2006 &                    **** &                      **** &                        ns \\
3  &             Singapore &        0.1326 &                    **** &                      **** &                        ns \\
4  &               Romania &        0.0946 &                    **** &                      **** &                        ns \\
5  &             Indonesia &        0.0926 &                    **** &                       *** &                        ns \\
6  &               Estonia &        0.0901 &                    **** &                      **** &                        ns \\
7  &                 Chile &        0.0891 &                       * &                        ** &                        ns \\
8  &                 China &        0.0865 &                      ns &                        ns &                        ns \\
9  &  Emirates &        0.0762 &                       * &                        ns &                        ns \\
10 &               Ecuador &        0.0755 &                      ** &                        ** &                        ns \\
11 &             Lithuania &        0.0679 &                    **** &                      **** &                        ns \\
12 &                 India &        0.0658 &                    **** &                      **** &                        ns \\
13 &               Vietnam &        0.0648 &                      ** &                        ** &                        ns \\
14 &              Bulgaria &        0.0627 &                     *** &                       *** &                        ns \\
15 &               Hungary &        0.0600 &                    **** &                      **** &                        ns \\
16 &                Turkey &        0.0587 &                      ns &                      **** &                        ns \\
17 &                Russia &        0.0559 &                      ns &                      **** &                        ns \\
18 &                 Japan &        0.0547 &                    **** &                      **** &                        ns \\
19 &              Malaysia &        0.0538 &                     *** &                       *** &                        ns \\
20 &                 Egypt &        0.0519 &                    **** &                       *** &                        ns \\
21 &              Colombia &        0.0504 &                       * &                        ** &                        ns \\
22 &                Canada &        0.0494 &                     *** &                      **** &                        ns \\
23 &               Ukraine &        0.0476 &                    **** &                       *** &                        ns \\
24 &               Austria &        0.0475 &                       * &                        ** &                        ns \\
25 &              Slovakia &        0.0471 &                      ** &                       *** &                        ns \\
26 &           South Korea &        0.0470 &                    **** &                      **** &                        ns \\

 27 &               Finland &        0.0455 &                    **** &                      **** &                        ns \\

\end{tabular}
\begin{tabular}{|llrlll}
\toprule
{} &               country & KL & KS test & MW test  & MW test  \\
{} &                &  &                &              (M$>$F)&   (F$>$M)\\
\midrule

28 &              Thailand &        0.0444 &                    **** &                      **** &                        ns \\
29 &               Croatia &        0.0407 &                      ns &                         * &                        ns \\
30 &                Poland &        0.0400 &                      ns &                      **** &                        ns \\
31 &           Netherlands &        0.0333 &                      ns &                      **** &                        ns \\
32 &             Argentina &        0.0321 &                      ** &                        ** &                        ns \\
33 &                Mexico &        0.0315 &                    **** &                      **** &                        ns \\
34 &                Brazil &        0.0308 &                    **** &                      **** &                        ns \\
35 &         United States &        0.0306 &                    **** &                      **** &                        ns \\
36 &        Czech Republic &        0.0306 &                    **** &                      **** &                        ns \\
37 &                 Spain &        0.0305 &                      ns &                      **** &                        ns \\
38 &               Belgium &        0.0301 &                      ns &                         * &                        ns \\
39 &               Denmark &        0.0275 &                    **** &                      **** &                        ns \\
40 &          South Africa &        0.0268 &                     *** &                      **** &                        ns \\
41 &                Taiwan &        0.0247 &                    **** &                      **** &                        ns \\
42 &                 Italy &        0.0246 &                    **** &                      **** &                        ns \\
43 &               Germany &        0.0241 &                    **** &                      **** &                        ns \\
44 &                  Iran &        0.0226 &                      ** &                         * &                        ns \\
45 &             Hong Kong &        0.0221 &                    **** &                      **** &                        ns \\
46 &           Philippines &        0.0203 &                      ns &                        ns &                        ns \\
47 &        United Kingdom &        0.0188 &                    **** &                      **** &                        ns \\
48 &             Australia &        0.0162 &                      ns &                        ns &                        ns \\
49 &               Ireland &        0.0155 &                      ns &                        ns &                        ns \\
50 &           Switzerland &        0.0139 &                      ns &                         * &                        ns \\
51 &                Norway &        0.0136 &                      ns &                      **** &                        ns \\
52 &                France &        0.0124 &                    **** &                      **** &                        ns \\
53 &                Sweden &        0.0089 &                      ns &                      **** &                        ns \\
 &           
 &&                     &                       &                        \\

\end{tabular}

\end{scriptsize}
\caption{\textbf{Gender differences in size of containers.} Statistics concerning the difference between the distribution of the 2-level containers sizes for males and females users in countries with more than $1000$ users. We report the Kullback-Leibler divergence (KL), as well as the p-values associated to the Kolmogorov–Smirnov test (KS test), and the Mann–Whitney U test testing that a randomly selected value from the male population is greater than a randomly selected value from the female population (MW test, M$>$F) and vice-versa (MW test, F$>$M). We used the following notation for p-values:  p$>$ 0.05 (ns),  p$\leq$ 0.05 (*), p$\leq$ 0.01 (**), p$\leq$ 0.001 (***), p$\leq$ 0.0001 (****)\label{TableGender2}}
\end{table}

\clearpage

\begin{table}
\begin{scriptsize}
\begin{tabular}{lrrlll}
\toprule
{} & level &      KL & KS test & MW test (urban$>$rural) & Mann–Whitney U test (rural$>$urban) \\
{} & \multicolumn{2}{l}{} & p-value &               p-value &                           p-value \\
\midrule
1 &       2 &  0.1416 &    **** &                    ns &                              **** \\
2 &       3 &  0.0807 &    **** &                    ns &                              **** \\
3 &       4 &  0.0523 &    **** &                    ns &                              **** \\
4 &       5 &  0.0444 &    **** &                    ns &                              **** \\
5 &       6 &  0.0439 &    **** &                    ns &                              **** \\
6 &       7 &  0.0457 &    **** &                    ns &                              **** \\
7 &       8 &  0.0594 &      ns &                    ns &                              **** \\
\bottomrule
\end{tabular}
\end{scriptsize}
\caption{\textbf{Urban/rural differences in the size of containers.} Statistics concerning the difference between the distribution of the containers sizes for urban and rural users. We report the Kullback-Leibler divergence (KL), as well as the p-values associated to the Kolmogorov–Smirnov test (KS test), and the Mann–Whitney U test testing that a randomly selected value from the urban population is greater than a randomly selected value from the rural population (MW test, urban$>$rural) and vice-versa (MW test, rural$>$urban). We used the following notation for p-values:  p$>$ 0.05 (ns),  p$\leq$ 0.05 (*), p$\leq$ 0.01 (**), p$\leq$ 0.001 (***), p$\leq$ 0.0001 (****)\label{TableUrban}}
\end{table}

\section*{Extended Data}

\setcounter{figure}{0} 
\setcounter{table}{0} 
  
\renewcommand{\figurename}{Extended Data Figure}
\renewcommand{\tablename}{Extended Data Table}

\begin{figure}[!htb]
    \includegraphics[width=\textwidth]{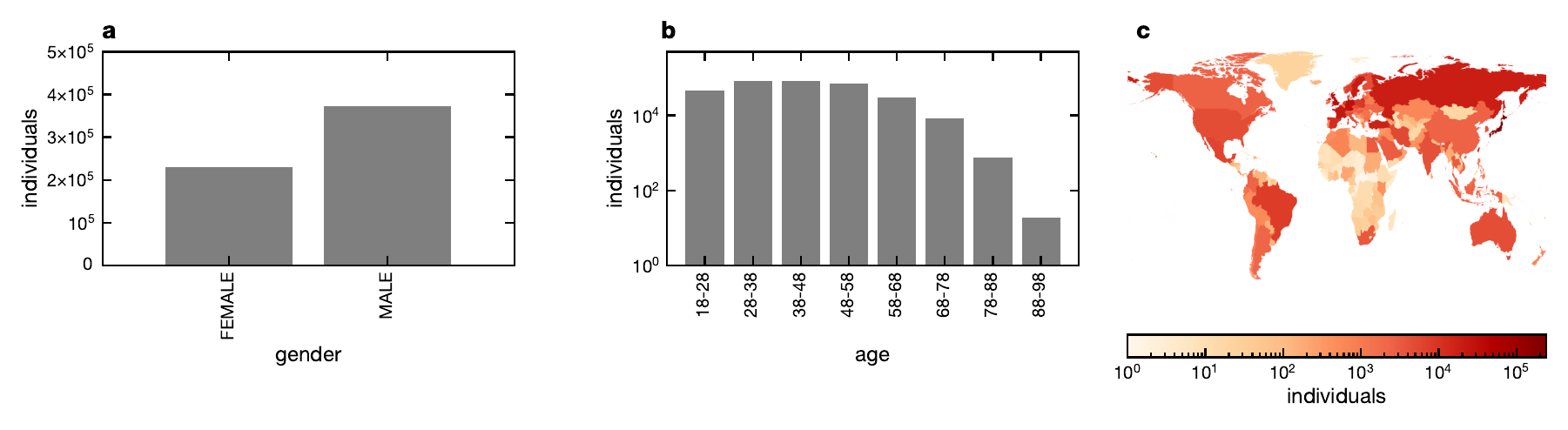}
    \caption{\textbf{The D1 dataset. }(a) Number of individuals for each gender. (b) Number of individuals per age group. (c) Number of individuals per country (see colorbar). We considered the $600,817$ individuals in our dataset with at least one year of data, and whose time-coverage (the fraction of time an individual position is known) was higher than $50\%$ at any given day. For these individuals, we considered one year of data with highest median time-coverage. Map data from the GADM Database of Global Administrative Areas, version 3.6, available at \href{www.gadm.org}{www.gadm.org}. \label{DataDescription}}
    \label{Extended_data_F1}
\end{figure}

\begin{table}[!htb]
    \includegraphics[width=\textwidth]{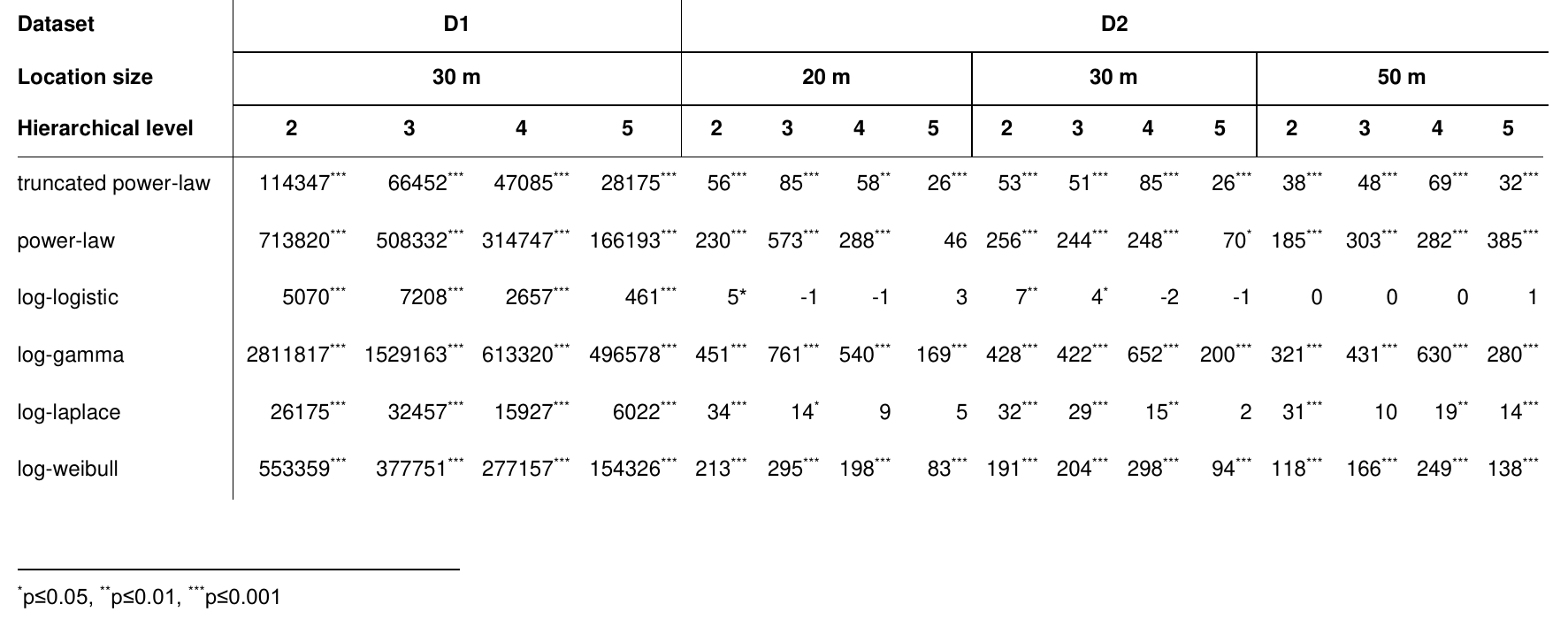}
    \caption{\textbf{The distribution of container sizes is not scale-free.} The log-likelihood ratio $R$\cite{clauset2009power} comparing the log-normal to other distributions (one per row) as a model for the distribution of container sizes. When $R$ is positive, the log-normal distribution has higher likelihood compared to the alternative, and vice-versa. The table reports also the p-values associated with $R$ (*$p\leq 0.05$,**$p\leq 0.01$,***$p\leq 0.001$). Results are shown at different hierarchical levels (rows), for dataset D2 and for dataset D1 under different choices of the parameter characterizing the typical size of individual locations.\cite{aslak2020infostop}}
    \label{Extended_data_T1}
\end{table}

\begin{figure}[!htb]
\centering
    \includegraphics[width=\textwidth]{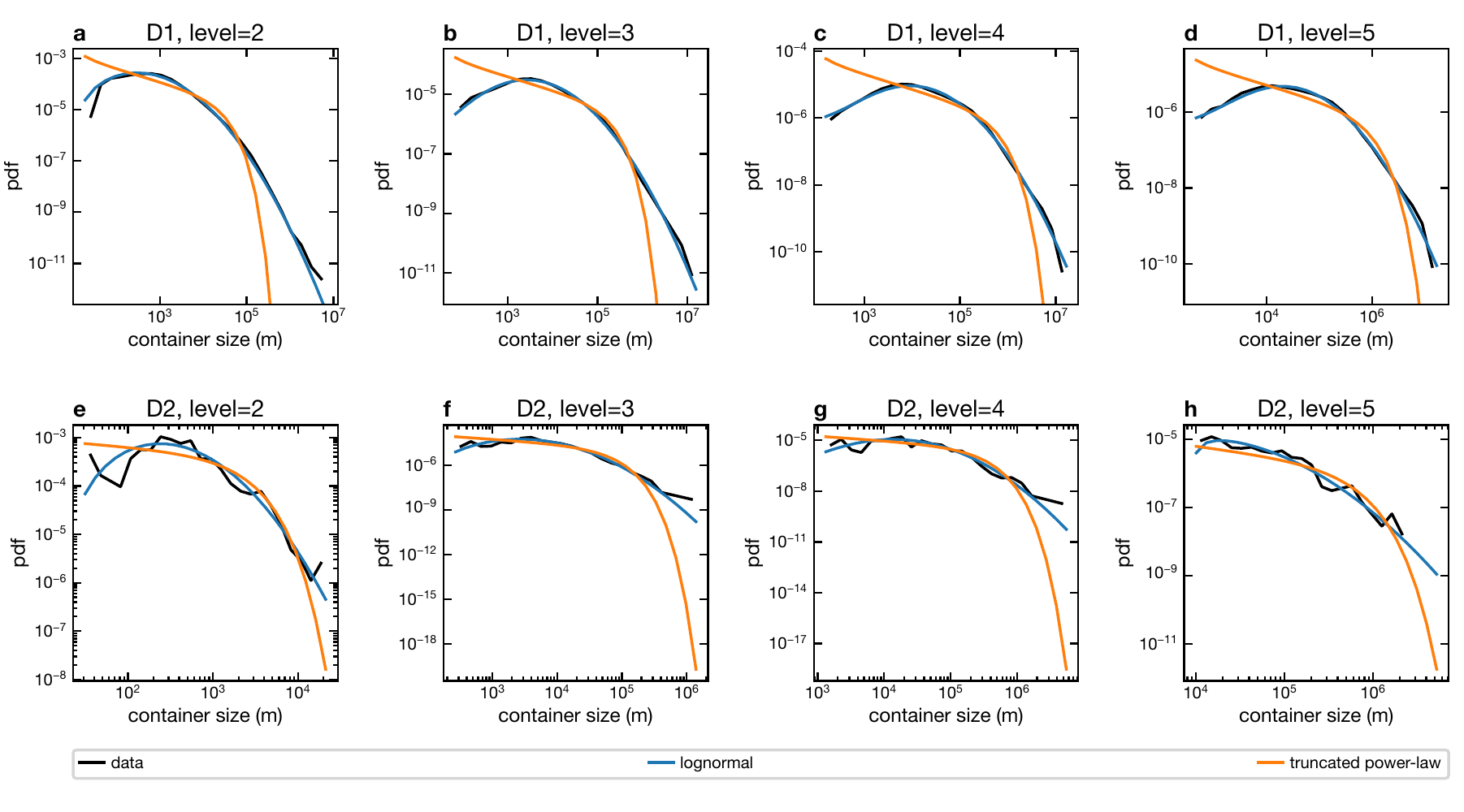}
    \caption{\textbf{Distribution of container sizes at different levels.} Distribution of individual container sizes at hierarchical levels $2$ (a,e), $3$ (b,f), $4$ (c,g), and $5$ (d,h), black line, and the corresponding lognormal (blue line) and truncated power-law (orange line) fits. Results are shown for the D1 (a to d) and D2 (e to h) datasets.}
    \label{Extended_data_F2}
\end{figure}

\begin{table}[!htb]
    \includegraphics[width=\textwidth]{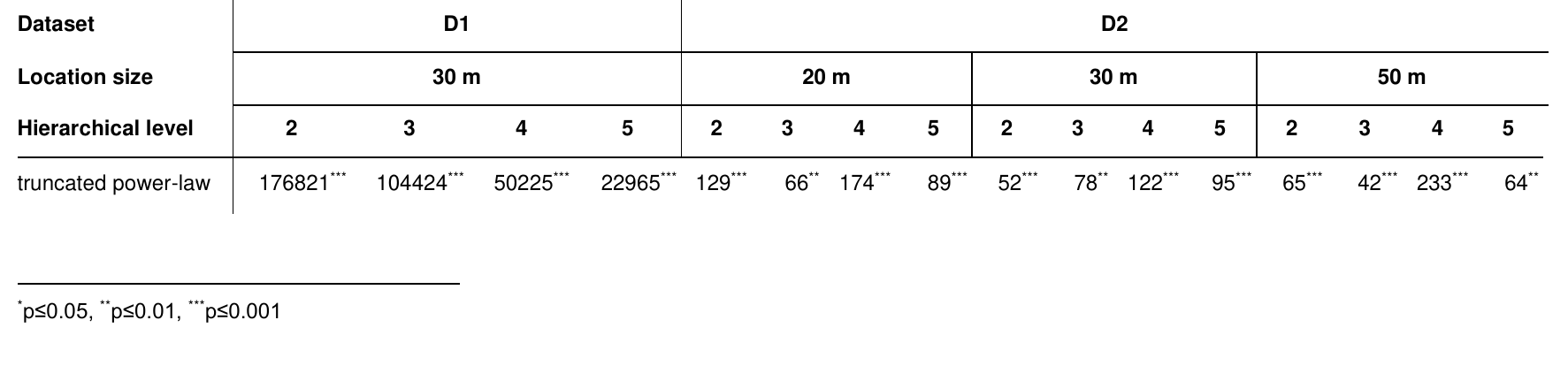}
    \caption{\textbf{The distribution of time spent within container is not scale-free.} The log-likelihood ratio $R$\cite{clauset2009power} comparing the log-normal to the truncated power law distribution as a model for the distribution of time spent within a container before transitioning to a different one. When $R$ is positive, the log-normal distribution has higher likelihood compared to the alternative, and vice-versa. The table reports also the p-values associated with $R$ (*$p\leq 0.05$,**$p\leq 0.01$,***$p\leq 0.001$). Results are shown at different hierarchical levels (rows), for dataset D2 and for dataset D1 under different choices of the parameter characterizing the typical size of individual locations.\cite{aslak2020infostop}}
    \label{Extended_data_T2}

\end{table}

\begin{table}[!htb]
    \centering
    \includegraphics[width=.5\textwidth]{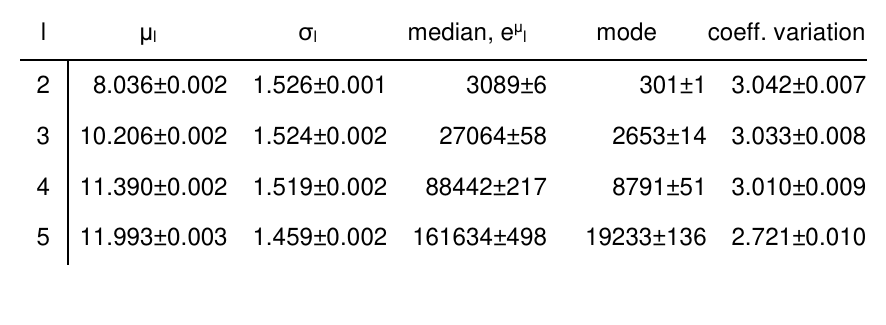}
    \caption{\textbf{Characteristics of the lognormal distributions of container sizes.} The parameters $\mu_l$ and $\sigma_l$ characterizing the lognormal distributions of container sizes at level $l$. We report also the median $e^{\mu_l}$, the mode, and the coefficient of variation $\sqrt{e^{\sigma_l^2}-1}$ defined as the fraction between the standard deviation and the mean\cite{romeo2003broad}}\label{char_lognorm}
    \label{Extended_data_T3}

\end{table}

\begin{figure}[!htb]
    \centering
    \includegraphics[width=\textwidth]{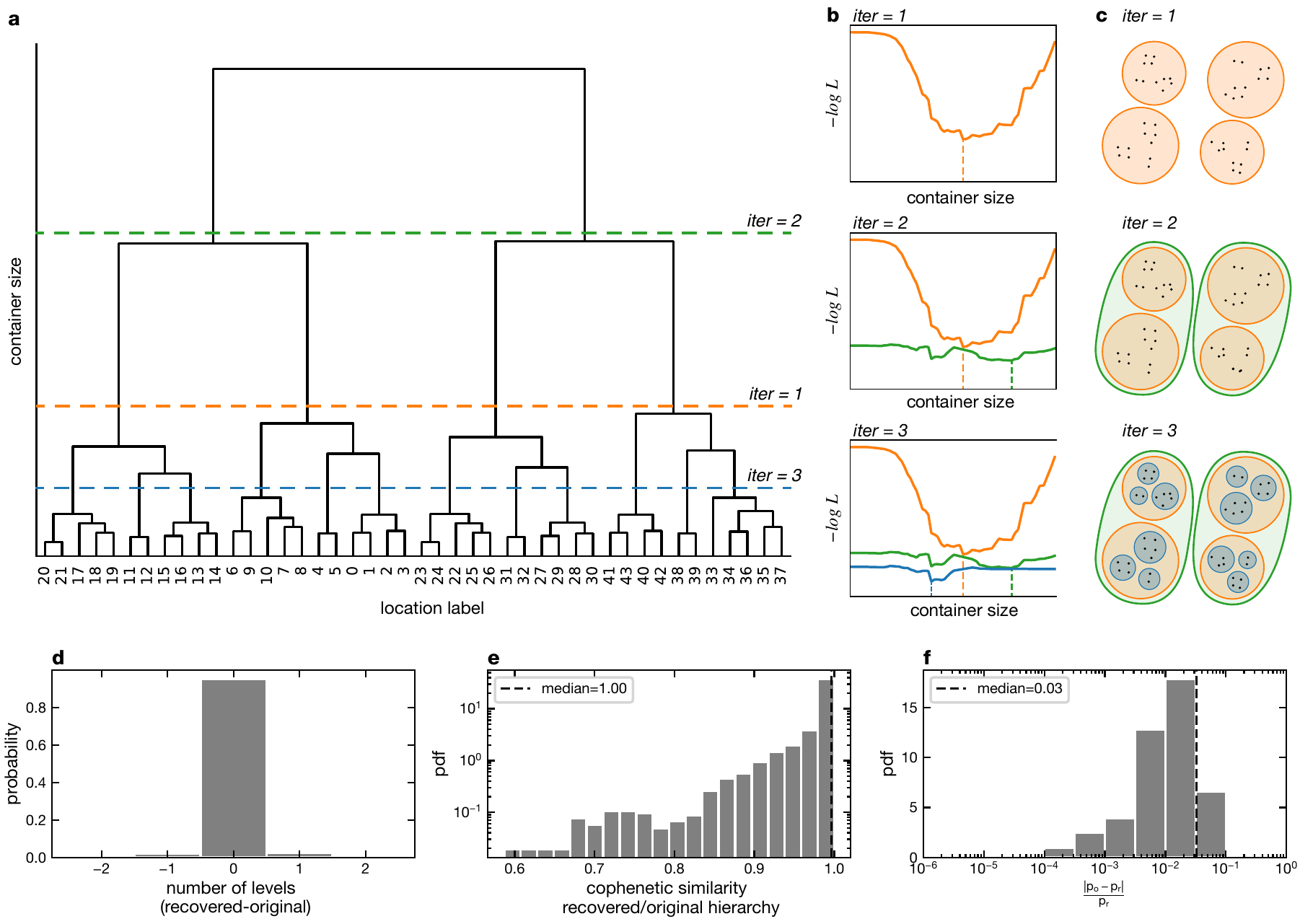}
    \caption{\textbf{Likelihood optimization: schematic description and validation.} We find the hierarchical partitioning corresponding to a sequence of locations as follows. (a) Individual locations are iteratively merged to form clusters via the complete linkage algorithm. Here, the output of the algorithm is visualized as a dendogram. (b) We add levels to the hierarchical partition by maximizing the likelihood of the container model \emph{one-level-at-a-time}: At the first iteration (top panel), we find the container size ($x$-axis) corresponding to the dendogram cut (dashed line) that minimizes the negative likelihood ($y$-axis), if any. We proceed by adding more dendogram cuts (central and bottom panels), and thus hierarchical levels, until the likelihood can not be further improved. (c) The dendogram cuts correspond to a hierarchical partitioning of individual locations. We evaluate the ability of the algorithm to recover the original parameters using $5,000$ synthetic traces of $3,000$ locations: (d) Distribution of the difference between the number of recovered and original levels. The difference is $0$ in $70\%$ of the cases. (e) Probability density associated to the cophenetic similarity between the original and recovered hierarchical structure. The dashed line corresponds to the median value. (f) Probability density associated to the relative difference $|p_{o} p_{r}|/p_{r}$ between original ($p_{o}$) and recovered ($p_{r}$) entries of the matrix $\boldsymbol{p}$.}
    \label{Extended_data_F3}
\end{figure}

\begin{figure}[!htb]
    \centering
    \includegraphics[width=\textwidth]{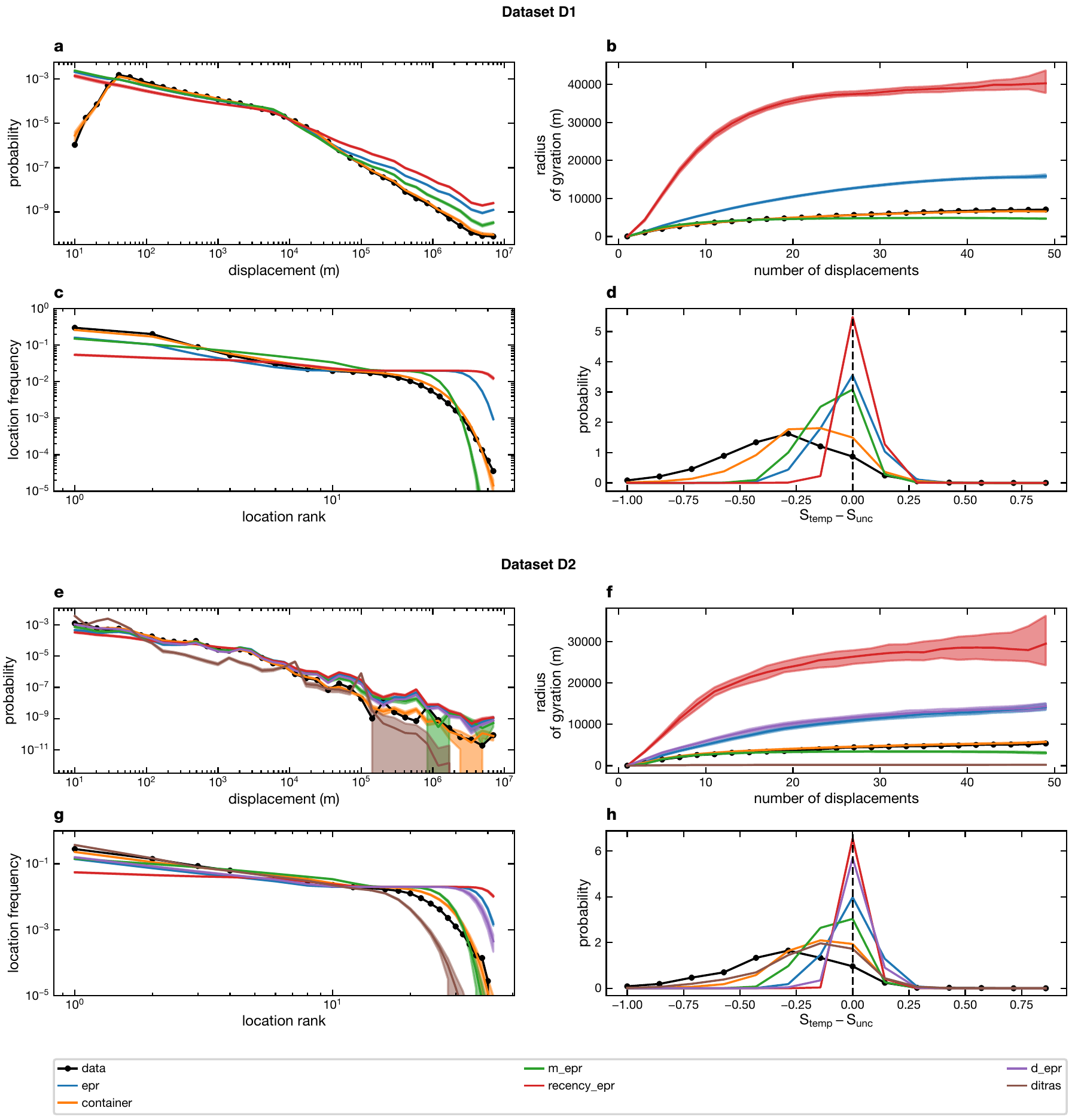}
    \caption{\textbf{The container model generate realistic synthetic traces.} (a-e) The distribution of displacements for the entire population, (b-f) The median individual radius of gyration vs the number of displacements, (c-g) The average visitation frequency vs the rank of individuals' locations, and (d-h) The distribution of the difference between the real entropy $S_{temp}$ and the uncorrelated entropy $S_{unc}$ across individuals. Results are shown for real traces (black line, dots), and traces generated by various models (see legend), for dataset D1 (a to d) and D2 (e to h). In panels a,c,d,e,g,h the filled areas for synthetic traces include two standard deviations around the mean computed across $1000$ simulations for each user. In panels b and f filled areas include the interquartile range. For each individual, we fitted the models considering a training period of $1$ year. The data used here for validation corresponds to the $50$ individual displacements following the training period.}
    \label{Extended_data_F4}

\end{figure}

\begin{table}[!htb]
    \centering
     \includegraphics[width=.5\textwidth]{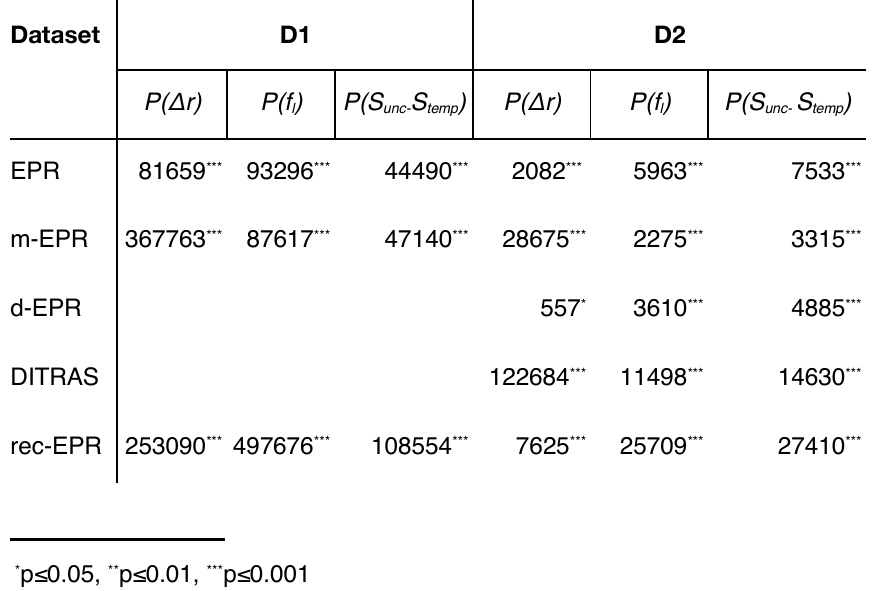}
   \caption{\textbf{The container model describes unseen data better than other individual mobility models.} The log-likelihood ratio $R$\cite{clauset2009power} comparing the likelihood of the container model to other models (one per row). When $R$ is positive, the container model has higher likelihood compared to the alternative, and vice-versa. The table reports also the p-values associated with $R$ (*$p\leq 0.05$,**$p\leq 0.01$,***$p\leq 0.001$). Results are shown for the D1 and D2 datasets. We report the results obtained considering different properties of the trajectories: the probability of displacement length $P(\Delta r)$, the probability of number of visits per location $P(f_l)$, the probability of the difference between the uncorrelated and temporal entropy $P(S_{unc} - S_{temp})$.}
    \label{Extended_data_T4}

\end{table}

\begin{figure}
\centering
    \includegraphics[width=.5\textwidth]{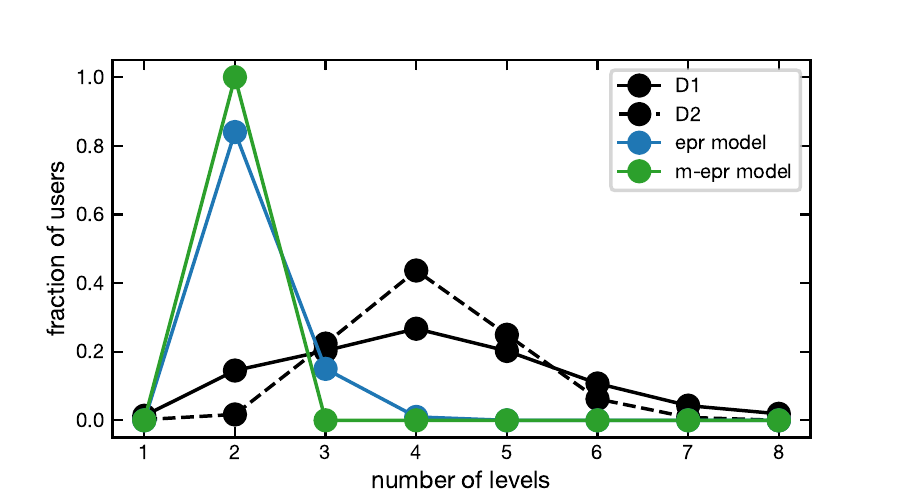}
    \caption{\textbf{Number of hierarchical levels recovered from traces.} Distribution of the number of hierarchical levels found by the container model for trajectories in the D1 dataset (plain black line), the D2 dataset (dashed black line), 1000 synthetic traces generated by the EPR model\cite{song2014prediction} (blue line) and m-EPR (green line) models. \label{Extended_data_F5}}
\end{figure}

\end{document}